\documentclass[10pt]{article}
\usepackage{bm}
\usepackage{graphicx}
\usepackage{amsmath}
\usepackage{amssymb}
\usepackage{amstext}
\usepackage{wrapfig}
\usepackage{color}
\begin{document} 
\title{Towards a unified linear kinetic transport model with the trace ion module for EIRENE}  
%
%
%
%
%
\author{J. Seebacher$^1$, A. Kendl$^1$\\
\vspace{0.1cm}
\footnotesize\it $^1$ Institute for Ion and Applied Physics, University of Innsbruck, \\
\footnotesize\it Association EURATOM-\"OAW, 6020 Innsbruck, Austria}

%
\date{}
\maketitle
\begin{abstract}
Linear kinetic Monte Carlo particle transport models are frequently employed in fusion plasma simulations to quantify atomic and surface effects on the main plasma
flow dynamics. Separate codes are used for transport of neutral particles (incl. radiation) and charged particles (trace impurity ions). Integration
of both modules into main plasma fluid solvers provides then self consistent solutions, in principle. The required interfaces are far from trivial, because
rapid atomic processes in particular in the edge region of fusion plasmas require either smoothing and resampling, or frequent transfer of particles from one into the other Monte Carlo code. We propose a different scheme here, in which despite the inherently  different mathematical form of kinetic equations
for ions and neutrals (e.g. Fokker-Planck vs. Boltzmann collision integrals) both types of particle orbits can be integrated into one single code.
We show that the approximations and shortcomings  of this ``single sourcing'' concept (e.g., restriction to explicit ion drift orbit integration) can be fully tolerable
in a wide range of typical fusion edge plasma conditions, and be overcompensated by the  code-system  simplicity, as well as by inherently ensured consistency in geometry (one single numerical grid only) and (the common) atomic and surface process modules.
\\

This is a pre peer reviewed version which has been submitted to Computer Physics Communications
\end{abstract}
\section{Introduction} \label{chaper_introduction}
A common computational approach for scrape off layer plasmas of fusion experiments are combined packages which consist of two coupled codes: a fluid solver for the main plasma (deuterium and tritium) and a kinetic solver for neutrals (deuterium, tritium, tungsten, carbon...), see e.g. \cite{Schn,EDGE2D,UEDGE,DEGAS}. In addition a kinetic code for impurity ions (tungsten, carbon, beryllium...) can be added, e.g. \cite{SOLDOR}. This approach avoids improper assumptions of the fluid modeling of impurity ions, which are instantaneous thermalization of the impurity particles and the lack of kinetic effects as a whole. Many standalone kinetic Monte Carlo impurity transport codes have been developed in the past, e.g. \cite{Stan,Sipi_Phd,Sipi,Rei1}. In general coupling of codes requires averaging of involved exchanged quantities due to different coordinate systems and grids used by the codes resulting in additional inaccuracies which degrade the output. Thus direct integration and streamlining of codes is favorable. The present approach combines kinetic treatment of neutrals with the kinetic treatment of impurity ions in linear approximation, where the plasma background is fixed. The kinetic neutral transport code EIRENE is supplemented by a \textit{trace ion module} (TIM), which comprises the numerical methods for solving a linear drift kinetic equation with Monte Carlo approach. 
The main difference of the TIM to other impurity codes is that it is fully part of EIRENE and many routines for describing the neutral particle dynamics are reused for treating charged particles. Thus EIRENE is now capable of solving three types of equations in one single code: the Boltzmann Equation for the neutrals, \cite{Reit,EIRE}, a Boltzmann like equation describing photon gas transport problems, \cite{Sven_Phd}, and TIM, which now allows solving the linear drift kinetic equation for impurity ions.

This paper is organized as follows: The current status of the Monte Carlo transport code EIRENE is summarized in the next chapter. Then the basic idea of the trace ion module is given in chapter 3. The linear drift kinetic model is shortly summarized in chapter 4 and its numerical implementation in the EIRENE code is described in chapter 5. The scope of application of the present trajectory integration method is analyzed in chapter 6 and the collision operator is verified in chapter 7. Finally the extended code is applied to a realistic modeling scenario for the MAST tokamak, where basic physics features of the trace ion module are verified. In addition the simulation results have been  compared with CCD camera images for two different wave lengths. The summary is given in chapter 7. Technical details are described in the appendix.
\section{Overview of the Monte Carlo Code EIRENE} \label{section_theory_part}
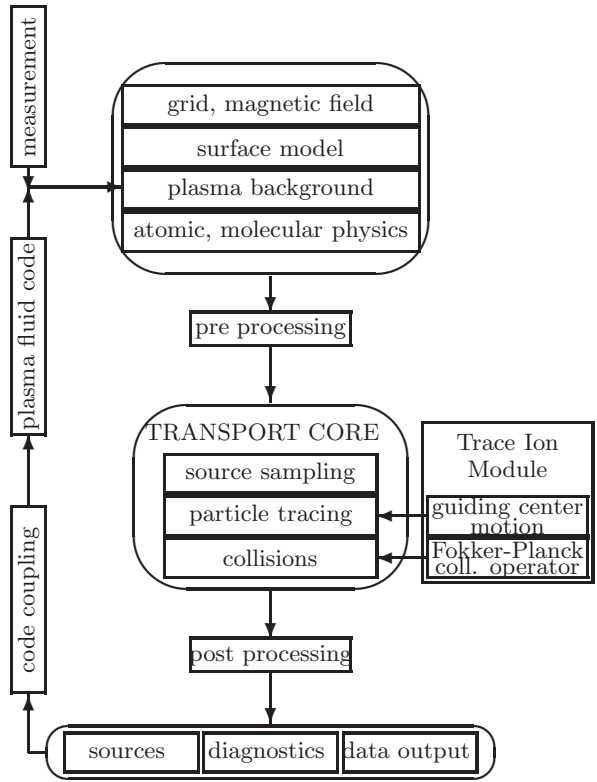
\begin{figure}
\setlength{\unitlength}{0.7cm}
\begin{center}
\begin{picture}(11 ,15) \thicklines

\put(5,0.85){\oval(8.5,1)}
\put(1.1,0.5){\framebox(2.5,0.7){\begin{small}sources \end{small}}}
\put(3.75,0.5){\framebox(2.5,0.7){\begin{small}diagnostics \end{small}}}
\put(6.4,0.5){\framebox(2.5,0.7){\begin{small}data output \end{small}}}

\put(5,2.4){\vector(0,-1){1}}

\put(3.5,2.4){\framebox(3,0.6){\begin{small} post processing \end{small}}}

\put(5,3.9){\vector(0,-1){0.9}}

\put(5,5.7){\oval(5.2,3.5)}
\put(3,5.8){\framebox(4,0.7){\begin{small} source sampling \end{small}}}
\put(3,5){\framebox(4,0.7){\begin{small} particle tracing \end{small}}}
\put(3,4.2){\framebox(4,0.7){\begin{small} collisions \end{small}}}
\put(2.5,6.6){\makebox(5,0.7){\begin{small}TRANSPORT CORE \end{small} }}

\put(8,5.35){\vector(-1,0){1}}
\put(8,4.55){\vector(-1,0){1}}

\put(7.9,4.1){\framebox(3.2,3.0){}}
\put(8,5){\framebox(3,0.7){}}

\put(8,5.35){\makebox(3,0.3){\begin{small}guiding center\end{small}}}
\put(8,5){\makebox(3,0.3){\begin{small}motion\end{small}}}

\put(8,4.2){\framebox(3,0.7){}}

\put(8,4.55){\makebox(3,0.3){\begin{small}Fokker-Planck\end{small}}}
\put(8,4.2){\makebox(3,0.3){\begin{small}coll. operator\end{small}}}
 
\put(8,6.4){\makebox(3,0.7){\begin{small} Trace Ion \end{small}}}
\put(8,5.9){\makebox(3,0.7){\begin{small} Module \end{small}}}

\put(5,9.9){\vector(0,-1){0.7}}

\put(3.5,8.6){\framebox(3,0.6){\begin{small} pre processing \end{small}}}

\put(5,8.6){\vector(0,-1){1.1}}

\put(5,11.95){\oval(6,4)}
\put(2.2,10.4){\framebox(5.6,0.7){\begin{small} atomic, molecular physics \end{small}}}
\put(2.2,11.2){\framebox(5.6,0.7){\begin{small} plasma background \end{small}}}
\put(2.2,12.0){\framebox(5.6,0.7){\begin{small} surface model \end{small}}}
\put(2.2,12.8){\framebox(5.6,0.7){\begin{small} grid, magnetic field \end{small}}}

\put(0.7,0.85){\line(-1,0){0.3}}
\put(0.4,0.85){\vector(0,1){1.1}}

\put(0.4,5.5){\vector(0,1){1.4}}
\put(0.4,10.6){\vector(0,1){1.0}}
\put(0.4,12.0){\vector(0,-1){0.5}}
\put(0.4,11.6){\vector(1,0){1.8}}

\put(0.1,2){\rotatebox{90}{\framebox(3.5,0.6){\begin{small} code coupling \end{small}}}}
\put(0.1,6.9){\rotatebox{90}{\framebox(3.7,0.6){\begin{small} plasma fluid code \end{small}}}}
\put(0.1,12){\rotatebox{90}{\framebox(3,0.6){\begin{small} measurement \end{small}}}}
\end{picture}
\end{center}
\caption{Outline of the Monte Carlo code EIRENE and where the \textit{trace ion module} is interfacing this code}
\label{FIG_FLOWCHART} 
\end{figure}

A brief overview of the current status of the Monte Carlo transport code EIRENE \cite{Reit,EIRE} is given. This code package has been developed for modeling neutral gas and radiation transport problems in magnetically confined plasmas. It is a multi-species code solving simultaneously a system of time dependent or stationary linear kinetic transport equations of almost arbitrary complexity. Linear kinetic transport problems are characterized by a given background plasma distribution function $f_b(\textbf{v})$, which is usually reconstructed from fluid quantities. There are no restrictions concerning the geometric complexity of the problem, in general any discretization from 0-3 dimensional computational domains is possible. It has been mostly used together with fluid transport codes, where it supplies the necessary particle, momentum and energy sources from the neutral particle dynamics to the fluid codes. In return these fluid codes provide the plasma background for EIRENE. 
Surface processes (physical and chemical sputtering as well as reflection) are treated by EIRENE with sputter models and reflection data bases, see \cite{ECK2}. Atomic and molecular data is obtained from external data bases, most notably ADAS \cite{Summ}. EIRENE further has access to the HYDKIN cross section data base for hydrocarbon molecules, Refs. \cite{Meth,Etha}, for simulation of the complicated catabolism mechanisms of these hydrocarbons in the fusion plasma. Since EIRENE solves a general kinetic equation radiation transport problems (photon gas simulations) can be treated as well, one such example is modeling of High-Intensity Discharge (HID) lamps with EIRENE \cite{Boern}. The code structure of the EIRENE code and possibly other Monte Carlo codes for solving linear transport problems is outlined in figure \ref{FIG_FLOWCHART}. 
An approximate model for transport of ionized particles along magnetic field lines has previously also been part of the EIRENE code. This model is now expanded by the \textit{trace ion module}, which treats kinetic transport physics of ions with less simplifications.
\section{Basic Idea of the Trace Ion Module}
The \textit{trace ion module} is an extension of the EIRENE code for solving a linear drift kinetic equation, see e.g. \cite{Hela}, for describing transport of charged particles in the existing code framework. The idea was to just add new orbit following routines for ions as well as a Fokker-Planck collision operator. In contrast to the neutral particle dynamics which is governed by the Boltzmann equation the drift kinetic equation is of Fokker-Planck type. Boltzmann equations describe discontinuous jump processes, where short range interactions between particles are dominant. On the other hand Fokker-Planck equations describe diffusion processes. The sample paths are described by stochastic ordinary differential equations, referred to as Langevin equations. Random sampling as well as time discretization is required for numerically solving stochastic ordinary differential equations. For more details on stochastic ordinary differential equations see e.g. \cite{CMPP,Kloe}. 

A combination of discontinuous jump processes and time discrete integration of stochastic ordinary differential equations is possible whenever the time discrete integration algorithm can be formulated as a jump process. In this case the Langevin equations which correspond to the drift kinetic equation have been solved by splitting the particle motion into a deterministic part which is interrupted by an artificial collision event. This collision event takes into account the effective action of the Coulomb force on the particle trajectory during one time step as well as the change of the velocity vector due to the electromagnetic forces in guiding center approximation. The major problem is that EIRENE is working with distances rather than time steps for calculating trajectories of the neutrals. In fact EIRENE determines the flight distance of a neutral particle from an effective mean free path and samples the type of collision at the point of collision from a discrete distribution of probabilities of the involved collision processes. Only once this distance and the point of collision is known a time step can be calculated, which is simply the amount of time necessary to reach the point of collision at the current particle speed.\\
The natural coordinate system for following of guiding center orbits is aligned with the magnetic field, which decouples parallel and perpendicular motion and significantly simplifies trajectory integration. This is conflicting with the neutral particle dynamics which is treated in Cartesian coordinates. In practice the EIRENE code is working with unstructured grids in a Cartesian coordinate system and the integration of the guiding center equations has to fit into this scheme. An obvious choice for orbit integration are Runge Kutta methods, which are frequently used in many kinetic plasma transport codes. In principle such methods could be incorporated into EIRENE under the expense of additional particle tracking on the grid. If such methods are applicable in EIRENE without a significant performance loss is to be investigated. Computationally less demanding but highly accurate methods for orbit integration are Adams Bashforth backward methods. Currently a 4-step Adams-Bashforth backward formula has been incorporated into EIRENE, which makes the calculation of the effective guiding center velocity simple and efficient. Moreover numerical errors due to Cartesian coordinates have been mitigated by significantly improving the interpolation and differentiation techniques on the numerical grid. Interpolation methods from finite element theory have been applied, where grid cells possess shape functions allowing to interpolate and differentiate electric and magnetic field vectors locally in each grid cell. Currently three types of cells are supported: 3-node triangles, 4-node quadrangles and 4-node tetrahedrons.

Coulomb collisions are treated as field collisions, where test particles interaction with the background plasma is described by evaluating Trubnikov/Rosenbluth potentials \cite{Trub}. This is done one the fly for non-Maxwellian distribution functions, which enables thermal force effects on a kinetic level. Since this collision operator in guiding center coordinates is singular an implicit method has been developed, which handles the singularity occurring in the perpendicular drift coefficient. The present approach, the details of which are summarized in \ref{A_IMPL_METH}, shows a solution for overcoming this problem in the framework of a Monte Carlo approach, while solutions for the same problem in context with finite difference methods have already been reported in e.g. \cite{ZAIT}. The location of the new orbit integration and collision routines of the \textit{trace ion module} in the code structure of EIRENE is sketched in figure \ref{FIG_FLOWCHART}. 
\section{Some remarks on the linear Drift \\Kinetic Model}
A summary of the drift kinetic model is given. Drift kinetic theory applies whenever the Larmor radius of the considered ion is much smaller than the gradient length of the external magnetic fields, $\rho_L\ll B/|\nabla B|$. The Larmor radius for ions with mass $m$ and charge number $Z$ is defined by $\rho_L=v_2/\Omega$, where $v_2$ is the perpendicular velocity component and the cyclotron frequency is given by $\Omega=ZeB/m$. The magnetic field vector is denoted by $\textbf{B}$, its absolute value by $B$ and the unit vector by $\textbf{b}$. The charge of the electron is $e$. The following notations are applied, 
\begin{eqnarray*}
	&v_1 = \textbf{v}\cdot \textbf{b}, &E_{\parallel}=\textbf{E}\cdot\textbf{b},\\ &v_2 = \sqrt{v^2-v_1^2},  &\nabla_{\parallel}=\textbf{b}\cdot\nabla, 
\end{eqnarray*} 
where the velocity component parallel to the magnetic field is denoted with $v_1$. The drift kinetic equation for stationary magnetic and electric fields and a fixed plasma background (linear approximation) can then be written as
\begin{eqnarray}
     \frac{\partial}{\partial t}\left(v_2f\right) = - \nabla_{\textbf{y}}(\dot{\textbf{y}}\cdot v_2f)  -\sum_{k=1}^2\partial_k\left([\dot{v}_k+A_k]v_2f\right)\nonumber\\  +\frac{1}{2}\sum_{k,l=1}^2\partial_k\partial_l\left(D_{kl}v_2f\right)+S, \label{drift_kin_equation}
\end{eqnarray}
which governs the time evolution of the distribution function $f(t,\textbf{y},v_1,v_2)$. For a detailed derivation of this equation see e.g. \cite{Hela}. This equation is already written in Fokker-Planck form which allows straight forward construction of appropriate Monte Carlo methods. Therein the guiding center velocity $\dot{\textbf{y}}$ and the time derivatives of the reduced phase space velocities $v_1,\,v_2$ are defined by
\begin{eqnarray}
   \dot{\textbf{y}}&=&v_1\textbf{b}+\textbf{v}_D\nonumber\\
    \textbf{v}_D&=&\frac{\textbf{E}\times\textbf{B}}{B^2}+\frac{1}{2}\frac{v_2^2}{\Omega}\frac{\textbf{b}\times \nabla B}{B}\nonumber\\ &&+\frac{v_1^2}{\Omega}\textbf{b}\times\left(\textbf{b}\cdot\nabla\right)\textbf{b} \nonumber\\
    \dot{v}_1&=&\frac{Ze}{m}\textbf{E}_{\parallel}-\frac{1}{2}v_2^2\frac{\nabla_{\parallel} B}{B} \nonumber \\
    \dot{v}_2&=&\frac{1}{2}v_1v_2\frac{\nabla_{\parallel} B}{B} \label{equs_of_motion}
\end{eqnarray}
This approach covers the drifts due to radial electric field and magnetic field inhomogeneities as well as the mirror effect and the electrostatic force. Coulomb collisions are represented by the drift and diffusion coefficients, $A_k$ and $D_{kl}$ respectively. A  comprehensive report on evaluating these coefficients for non-Maxwellian plasmas  can be found in \cite{Rei2}.  Therein appropriate drift and diffusion coefficients in the reduced $(v_1,\,v_2)$ phase space are given,
\begin{eqnarray}
    A^1 &=& \mu \Lambda\frac{\partial\phi}{\partial v_1},\nonumber\\
    A^2 &=& \mu \Lambda\frac{\partial\phi}{\partial v_2}+\frac{1}{2}\Lambda\frac{1}{v^2_{2}}\frac{\partial\psi}{\partial v_{2}},\nonumber\\
    D^{11}&=&\Lambda\frac{\partial^2\psi}{\partial v^2_{1}},\nonumber\\
    D_{22}&=&\Lambda\frac{\partial^2\psi}{\partial v^2_{2}},\nonumber\\
    D^{12}&=&\Lambda\frac{\partial^2\psi}{\partial v_{1} \partial v_{2}}= D_{21},
     \label{diff_coeff_1}.
\end{eqnarray}\\
The Trubnikov/Rosenbluth potentials $\phi$ and $\psi$ are specialized for taking into account friction as well as the thermal force effect. The derivation of the explicit form of these potentials is given in \ref{A_13MINT}, which differs from the derivation given in \cite{Rei2} in the following respects: a different method has been used for integration of the potential functions and the differentials required for calculating the thermal force effect have been expanded to cover the perpendicular direction, which will enable thermal force effects in radial direction in the future. It is noted that the notations and definitions of the potentials derived in \cite{Rei2} have been reused in this work. The standard Monte Carlo approach for diffusion in real space with a constant diffusion coefficient is used, where the diffusion matrix is of the form
\begin{equation}
	D_{ij}=\partial^2/\partial y_i\partial y_j\left(e_i e_k D_{\perp}v_2f\right).
\end{equation}
Therein the vector $\textbf{e}$ is defined to be perpendicular to the magnetic field in radial direction lying in the poloidal plane. The factor $\Lambda$ is given by 
\begin{equation*}
    \Lambda=\lambda n_b\frac{Z^2 Z_b^2 e^4}{4\pi\epsilon_0^2 m^2}. 
\end{equation*}
where $n_b$ and $Z_b$ denotes particle density and charge of the background particles. The Coulomb logarithm $\lambda$ replaces a singular integral which results from the infinite range of the electrostatic potential. This integral is usually cut off at a certain impact parameter which is of the order of the Debye radius. The Coulomb logarithm has been taken constant through out the whole work. As a typical value for fusion relevant plasmas $\lambda=13.5$ has been used. Since impurities in the SOL plasma consist of many different species and each of these species can have more than one charge state additional sources/sinks represented by $S$ in equation \ref{drift_kin_equation} account for ionization, recombination $(S_I,S_R)$ and external particle creation and annihilation processes $(Q_Z^{\pm})$. The complete source/sink term reads 
\begin{eqnarray*}
	S &=& -S_I^Z(v_2f)^Z+S_I^{Z-1}(v_2f)^{Z-1}\\&&-S_R^Z(v_2f)^Z+S_R^{Z+1}(v_2f)^{Z+1}
                                     +Q_Z^+-Q_Z^-. 	
\end{eqnarray*}
%
\section{Numerical Implementation} \label{subsection_langevin}
Random walks of neutral particles in the EIRENE code are constructed by evaluating the distance $d$ to the next collision according to the effective local mean free path of the involved collision processes. Once this distance is known EIRENE updates the position of a particle according to 
\begin{equation}
           \textbf{y}_{new}=\textbf{y}_{old}+\textbf{e}_v\cdot d \label{advance_position}
\end{equation}
where particle positions are denoted by $\textbf{y}$ and the unit velocity vector is $\textbf{e}_v$. At the new position the next collisional event is executed and the velocity vector $v\cdot\textbf{e}_v$ of the particle is instantaneously changed. This process is continued until a time limit is reached or the particle is absorbed at a surface. Macroscopic quantities are estimated from the random walks of the particles on the computational grid by applying a track length estimator. Random walks of this type, referred to as Markov chains, solve the Boltzmann equation for the neutrals in terms of distribution functions.

In the present approach the dynamics of ionized particles is governed by stochastic ordinary differential equations corresponding to \ref{drift_kin_equation}. The first order Euler-Maruyama Method for discretization of these equations is given by
\begin{equation}
 	X_{t+\Delta t}=X_t+A(X_t)\Delta t+B(X_t)\sqrt{\Delta t}\zeta,
\end{equation}
where $X=\lbrace X_t,t\geq 0\rbrace$ is the state vector of the system, which is represented in this case by the position of the particle $\textbf{y}$ and the reduced $(v_1,v_2)$ velocity phase space. Local drift coefficients are denoted by $A\equiv(\textbf{y},v_{1,2}+A_{1,2})$, diffusion coefficients by $B$ and $\zeta$ is a normal distributed random number with mean $0$ and variance $1$. Diffusion coefficients $B$ can be constructed from the corresponding diffusion matrix ($D=BB^T$) in \ref{drift_kin_equation}, which in principal is a five dimensional matrix covering diffusion in real space as well as in velocity space. An appropriate method for constructing $B$ in velocity space is given in \ref{A_DRIFTDIFFCOEFF}. The initial state of a new born particle, due to e.g. a surface process, is the initial position and the parallel and perpendicular velocity components of the 3D velocity vector, which EIRENE assigns to newly created particles.

For incorporating the discretized Langevin equations into the Monte Carlo framework of the EIRENE code the actual algorithm has to be formulated as a jump process. In particular the position of the ionized particle has to be updated in the same way as for the neutrals according to eq. (\ref{advance_position}), but with an effective guiding center velocity 
\begin{equation*}
v\rightarrow\|\dot{\textbf{y}}\| \hspace{0.3cm}  \textbf{e}_v\rightarrow\dot{\textbf{y}}/\|\dot{\textbf{y}}\|.\nonumber
\end{equation*}
This effective guiding center velocity vector is constructed at each point of collision and takes into account the effect of the electromagnetic force as well as the Coulomb force both acting on the particle over a time period $\Delta t$. Currently a 4-step Adams Bashforth backward formula is used for evaluating $\dot{\textbf{y}}$, the details of which are given in \ref{A_ADAMS_BASHFORTH}. For calculating a distance to the next Coulomb collision event, 
\begin{equation}
d=v\cdot\Delta t_{CC}, 
\end{equation}
an appropriate time step $\Delta t_{CC}$ has to be supplied. This is not necessarily the time step used for the actual collision process, but it allows EIRENE to determine the shortest distance for the next type of collision. Coulomb collisions have to be executed in any case, no matter what other collision event is executed. In practice the time step for Coulomb collisions and furthermore the distance to the next Coulomb collision has to be much smaller than the mean free path of the other involved collision processes for proper resolving the thermalization of the test particles with the plasma background. In fact $\Delta t_{CC}$ is just the upper limit for the actual time step. The actual time step for updating the guiding center position can be calculated after the flight distance $d$ has been determined by EIRENE from  
\begin{equation}
\Delta t=d/\|\dot{\textbf{y}}\|<\Delta t_{CC}.
\end{equation}
For resolving thermalization processes $\Delta t_{CC}$ can be adjusted as a fraction, e.g. $1/100$, of the local equilibration time.

The actual algorithm works in the following manner: after the distance to the next collision has been determined by EIRENE and the time step $\Delta t$ is known the particle position is updated according to equation \ref{advance_position}, where $\textbf{e}_v\cdot d$ is replaced with $\Delta t\cdot\dot{\textbf{y}}(t)$. In the next step post collision phase space velocities are calculated by 
\begin{equation}
   \chi_{i}^{post-col}= \chi_{i}^{pre-col}+\Delta\chi^{det}_{i}+\Delta\chi^{stoch}_{i}, 
\end{equation}
where normalized quantities are used, $\chi_i=\alpha\,v_i$ with the inverse thermal velocity $\alpha=\sqrt{2\,T_b/m_b}$. This is an effective collision where in addition to the Coulomb interaction, $\Delta\chi^{stoch}$, also the acceleration due to the electric field and the mirror force, $\Delta \chi^{det}$, is accounted for. The latter are calculated at the current particle position $\textbf{y}_{post-col}$ and are given by
\begin{eqnarray}
    \Delta\chi^{det}_{1} &=& \alpha\left(\frac{Ze}{m}\textbf{E}_{\parallel}-\frac{1}{2}v_{2}^2\frac{\nabla_{\parallel}B}{B}\right)_{\textbf{y}=\textbf{y}_{post-col}}\Delta t \nonumber \\
    \Delta\chi^{det}_{2} &=&
    \alpha\left(\frac{1}{2}v_{1}v_{2}\frac{\nabla_{\parallel}B}{B}\right)_{\textbf{y}=\textbf{y}_{post-col}}\Delta t. 
\end{eqnarray}
It is noted that consistency requires to evaluate the deterministic velocity increments $\Delta\chi^{det}$ in the same way as $\dot{\textbf{y}}$. The stochastic velocity increments are evaluated in any case according to 
\begin{equation}
   \Delta\chi^{stoch}_{i}=A_{i}\Delta t\,+\sum_{j=1}^{2}B_{ij}\sqrt{\Delta t} \zeta_{j},
\end{equation}
with two random numbers $\zeta_{1,2}$. 

The singularity of this collision operator is expressed in the perpendicular drift coefficient $A_2$, which is singular for $\chi_{2}\rightarrow 0$. If friction forces only are considered a different set of coordinates might have allowed to get around this difficulty, but for a numerical implementation of the thermal force effect in the kinetic Monte Carlo approach the present coordinates $(v_1,v_2)$ are the most simple and computationally the most efficient ones. The present algorithm has been supplemented by an implicit part, which treats slow particles (with respect to $v_2$) below a certain threshold. The additional computational expense is about $10\%$, where it is noted that without this implicit part the energy of the test particles is overestimated and conservation of energy of the collision operator is severely violated. Details are described in \ref{A_IMPL_METH}.

The next step of this algorithm is finding interpolated values and derivatives of the magnetic and electric fields at the point of collision $\textbf{y}_{post-col}$, 
\begin{equation*}
 	\dot{\textbf{y}}\left(\chi_{1}^{post-col},\chi_{2}^{post-col},\textbf{B}(\textbf{y}_{post-col}),\textbf{E}(\textbf{y}_{post-col})\right).
\end{equation*}
and evaluating the new effective guiding center velocity by applying an appropriate integration method. Finally the particle position is advanced to the next point of collision. In this approach $\chi_{1}$ and $\chi_{2}$ are actually dummy variables which are only required for constructing the guiding center velocity. In fact an ionized particle is described in EIRENE by the position $\textbf{y}$ and the velocity $\dot{\textbf{y}}$ in the six dimensional 3D in real and 3D in (guiding center) velocity space.
%
It is noted that in 2D simulations the toroidal contribution of the curvature drift cannot be resolved. The magnetic field is given on the poloidal plane only, which is in fact a cylindrical approximation. A possible axis symmetric toroidal magnetic field component has the form  $\textbf{B}(r,\theta)=B_0R_0/R\hat{\textbf{e}}_{\varphi}$, where $\hat{\textbf{e}}_{\varphi}$ is the unit vector in toroidal direction. The toroidal part of the curvature drift is then recovered as
\begin{equation}
     v_{CD}=-\textbf{b}\times\left(\textbf{b}\times\left(\nabla\times\textbf{b}\right)\right)=-\frac{1}{R}\textbf{e}_y, \label{CD_quasi3d_correction}
\end{equation}
where $\textbf{e}_y$ is a unit vector pointing upward in the direction of the axis of the tokamak (the plane in which the torus is located is the $\textbf{e}_x\times\textbf{e}_z$-plane). The direction of this drift is aligned with the $\nabla B$ drift, which is assumed to point downwards here. Especially for spherical tokamaks, e.g. MAST, the toroidal curvature is one order of magnitude higher as for other tokamaks and the effect of the toroidal curvature has a big impact on modeling results.
\section{Guiding Center Orbit Integration}\label{subsection_orbit_integration}
\begin{figure}[htbp]
\begin{center}
\includegraphics[width=7.5cm]{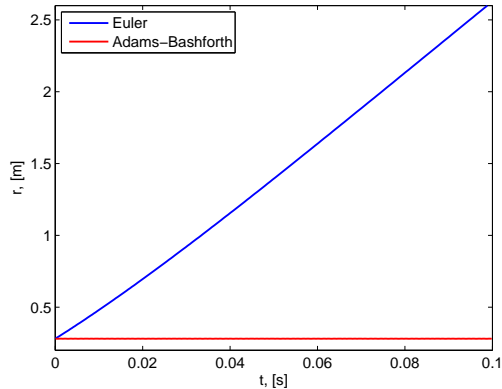}
\caption{Time evolution of radial coordinate at mid plane of closed banana orbit for simple first order orbit integration and 4-step Adams Bashforth backward method, magnetic field in circular approximation with dimensions of MAST closed field line region, collision less particle launched close to separatrix}
\label{FIG_Adams-Bashforth}
\end{center}
\end{figure}

The quality of orbit integration in the presented algorithm is determined by the evaluation of the effective guiding center velocity at each time step. If $\dot{\textbf{y}}$ is calculated from the local plasma and magnetic field parameters at the current simulation time the resulting jump process is identical to the first order Euler method and fits perfectly to EIRENE by lowest computational costs. The drawback of applying the Euler method for guiding center orbit integration in fusion plasmas is the accumulating numerical error, which leads to an artificial outward drift of the ions. This is only tolerable for short living particles where closed orbits are not to be expected. For example the life time of $C^{++}$ in the divertor region of the MAST tokamak is $~200\mu s$ according to ionization rates of ADAS, \cite{Summ}. This is long enough to fully thermalize. On the other hand the bounce time estimated for a circular magnetic field (major radius $R_0$ chosen for MAST edge plasma) and a particle at a thermal speed of $10^{4}m/s$ is $\tau_b=10^{-2}s$. The lifetime of $C^{++}$ is two orders of magnitude lower than the time for passing at least a single orbit, which justifies to use the simple Euler orbit integrator in this case.\\ 
For treating particle motion of particles with longer life times more accurate orbit integration methods have been sought. In fusion research Runge Kutta methods are frequently used for guiding center orbit following codes. One such example is ASCOT, \cite{Sipi_Phd,Sipi}. Implementing an $n-th$ order Runge Kutta methods into EIRENE requires the evaluation of $n$ intermediate steps, for which each time the whole geometry module (locate particle on grid, interpolate value inside grid cell) has to be called. Since the geometry module is the most demanding part of EIRENE with respect to computing power such methods are not suitable. Promising methods, which allows to improve the orbit integration of EIRENE while keeping the additional computational effort low, are Adams Bashforth backward formulas. The effective velocity vector needed for advancing the particle position is constructed from previous velocity values rather than from calculating intermediate steps, which requires almost no additional computing power just a negligible increase of storage.
The details of the 4-step backward method which has been incorporated into the EIRENE code is described in \ref{A_ADAMS_BASHFORTH}. The reason why a 4-step backward method has been used is because backward formulas taking two and three steps were observed to be unstable. Nevertheless these formulas are required to start the particle orbit. This unstable behavior might be due to extreme time step differences, which can occur in EIRENE. In particular in front of surfaces the time step can change e.g. from $\Delta t=10^{-7}$ to $\Delta t=10^{-10}$, because EIRENE is moving a particle from one cell to the cell face of the next one for keeping track of the particles on the grid. This method can be easily extended to 5 or more step method, but this has not been considered yet.\\
As an illustrative example of the performance of the backward orbit integrator is shown in figure \ref{FIG_Adams-Bashforth}. The evolution of the radial coordinate at the mid plane ($v_{\parallel}$ positive, outer zero crossing of the orbit) of a closed collision less orbit is plotted vs. time. The orbit, which has been calculated with the Euler method, continuously drifts outward while the radial coordinate of the orbit obtained with the 4 step Adams-Bashforth keeps constant (300 orbits transits are shown). The outward drift per orbit is not completely removed with the Adams-Bashforth method, but compared to the Euler integrator it is considerably lower. For this particular orbit, which has been integrated with a step size of $\Delta t = 10^{-7}s$, the radial outward drift per orbit has been found to be $dr_{\text{Adams}}/d\tau\sim 10^{-6}m/\tau \ll dr_{\text{Euler}}/d\tau\sim 10^{-3}m/\tau$, where $\tau$ denotes the amount of time to pass one full orbit.
%
\section{Verification of the Coulomb Collision Operator} \label{subsection_equilibration}
\begin{figure}[htbp]
\begin{center}
\includegraphics[width=7.5cm]{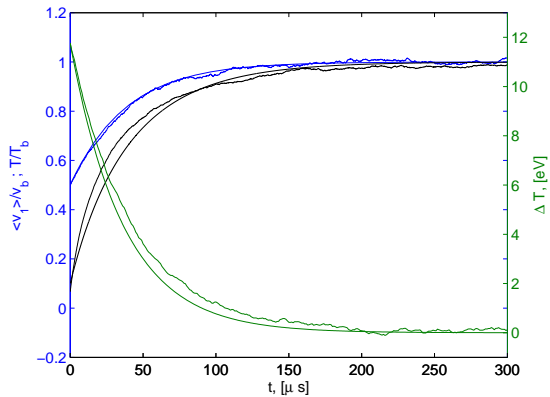}
\caption{Comparison of relaxation time approximations and Monte Carlo simulation, left ordinate corresponds to normalized $\langle v\rangle/u$ (slowing down, blue line) and $T/T_b$ (thermalization, black line) and the right ordinate corresponds to $\Delta T$ (isotropization, green line)}
\label{FIG_Relaxation_and_Grid_dependence}
\end{center}
\end{figure}

Whereas the steady state behavior of the present Cou- lomb collision operator has been well tested in Ref. \cite{Rei2}, verification of the dynamics of the collision operator is missing. For this reason the time evolution of corresponding moments of the kinetic distribution function have been estimated and the results have been confronted to available analytical relaxation time approximations. In particular, slowing down of particles, temperature equilibration and temperature isotropization have been checked. The most simple estimator has been used for calculating moments of the distribution function, $M$, which is given $M=1/(\Delta V_i N_i)\sum_i g(v_i)$, where $g(v_i)$ is any function of $v_i$, e.g. particle energy $m\,v_i^2/2$. A slab case for the EIRENE code has been set up existing of one cell with periodic boundary conditions and uniform magnetic field in z direction. Typical plasma parameters for fusion edge plasmas have been adjusted in these simulations, $T_b=10$eV, $n_{D+}=10^{18}$m$^{-3}$ and $\textbf{u}=0\,ms^{-1}$. A point source of test particles in the cell center has been placed, where $C^{2+}$ ions with isotropic, monoenergetic velocity distribution have been released at an initial energy of $1$eV. The time interval for the simulations was $1/2$ms, where $t_{initial}=0\,\mu s$ and $t_{final}=500\,\mu s$ and the time step has been fixed to $\Delta t=10^{-7}s$. An analytic expression for the equilibration of plasmas at different temperatures but no relative drift velocity is given by 
$dT/dt=\nu_{E}\left(T_b-T\right)$ where the characteristic frequency is 
\begin{equation}
\nu_{E}=\frac{4}{3\sqrt{2\,\pi}}\Lambda\frac{m^{5/2}}{m_b}\left(T+\frac{m}{m_b}T_b\right)^{-3/2}.
\end{equation}
Temperature isotropization of a plasma with constant overall temperature but different $T_{\parallel}$ and $T_{\perp}$ is described by $d\Delta T/dt=-\nu_{I}\Delta T$ where $\Delta T=T_1-T_2$ and the characteristic frequency
\begin{equation}
\nu_{I}=\frac{3}{2\sqrt{\pi}}\Lambda\left(\frac{m}{T_{\parallel 0}}\right)^{3/2}\gamma.
\end{equation}
where $\gamma=1/A^2\left(-3+(3-A)\,\text{atanh}\,(\sqrt{A})/\sqrt{A}\right)$ and $A=1-T_{\parallel 0}/T_{\perp 0}$. In this case particles have to be sampled from a bi-Maxwellian distribution function. The overall temperature, $T=(T_{\parallel}+2T_{\perp})/3$, is constant in this case and energy is solely transferred from parallel to perpendicular direction. Slowing down means that the macroscopic flow velocity of the test particles adjusts to the flow velocity of the plasma background, described by a shifted Maxwellian. The governing relaxation equation is $d\langle v\rangle=-\nu_S(\langle v\rangle - u_b)$ where the slowing down frequency is 

\begin{equation}
\nu_S=-\mu\Lambda\alpha^2\phi_0'(\chi)/\chi.
\end{equation}
The slowing down of test particles has been checked for a beam of particles aligned with the magnetic field. Results of the Monte Carlo simulations as well as the solutions of the ordinary differential equations for slowing down, thermalization and isotropization have been plotted in figure \ref{FIG_Relaxation_and_Grid_dependence}. 
Therein the time evolution of T and u, normalized to background plasma temperature as well as flow velocity, and the time evolution of $\Delta T$ is shown. Analytical results and simulation are in good agreement for all three processes.

Moreover the present collision operator includes the thermal force effect, the numerical implementation of which has been adapted for the EIRENE code but the net effect has already been described in \cite{Rei1,Rei2}. The main features are shortly summarized: The thermo effect is described in the present case on a kinetic level by taking into account a non equilibrium distribution function for the plasma background, denoted by $f_b$, in the derivation of appropriate Trubnikov/Rosenbluth potentials. In particular $f_b$ has been constructed from a Maxwellian distribution, which is perturbed by a series of Hermitian polynomials (see \ref{A_13MINT}). 
The coefficients in the perturbation series mainly depends on the temperature gradient and the ratio $T_b/n_b$. For high densities, $n_b\sim 10^{20}\,/m^3$, the influence of the temperature gradient is small, because the collisionality is high and the plasma is always close to local equilibrium. At low densities, e.g. closer to the target plates of a fusion device, even small temperature gradients lead to a significant temperature gradient force. Nevertheless the perturbation from equilibrium may not exceed the limit where $f_b<0$.
\section{A reference modeling scenario for the MAST tokamak} \label{section_mast_tokamak}
%
\begin{figure}[htbp]
\begin{center}
\includegraphics[width=7cm]{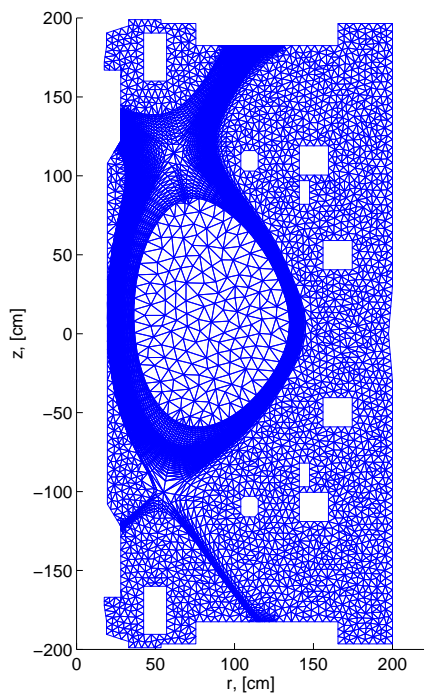}\\
\caption{EIRENE grid for MAST discharge \#$13949$}
\label{PIC_MAST_13949_GRID}
\end{center}
\end{figure}
%
\begin{figure}[htbp]
\begin{center}
\includegraphics[width=5.3cm]{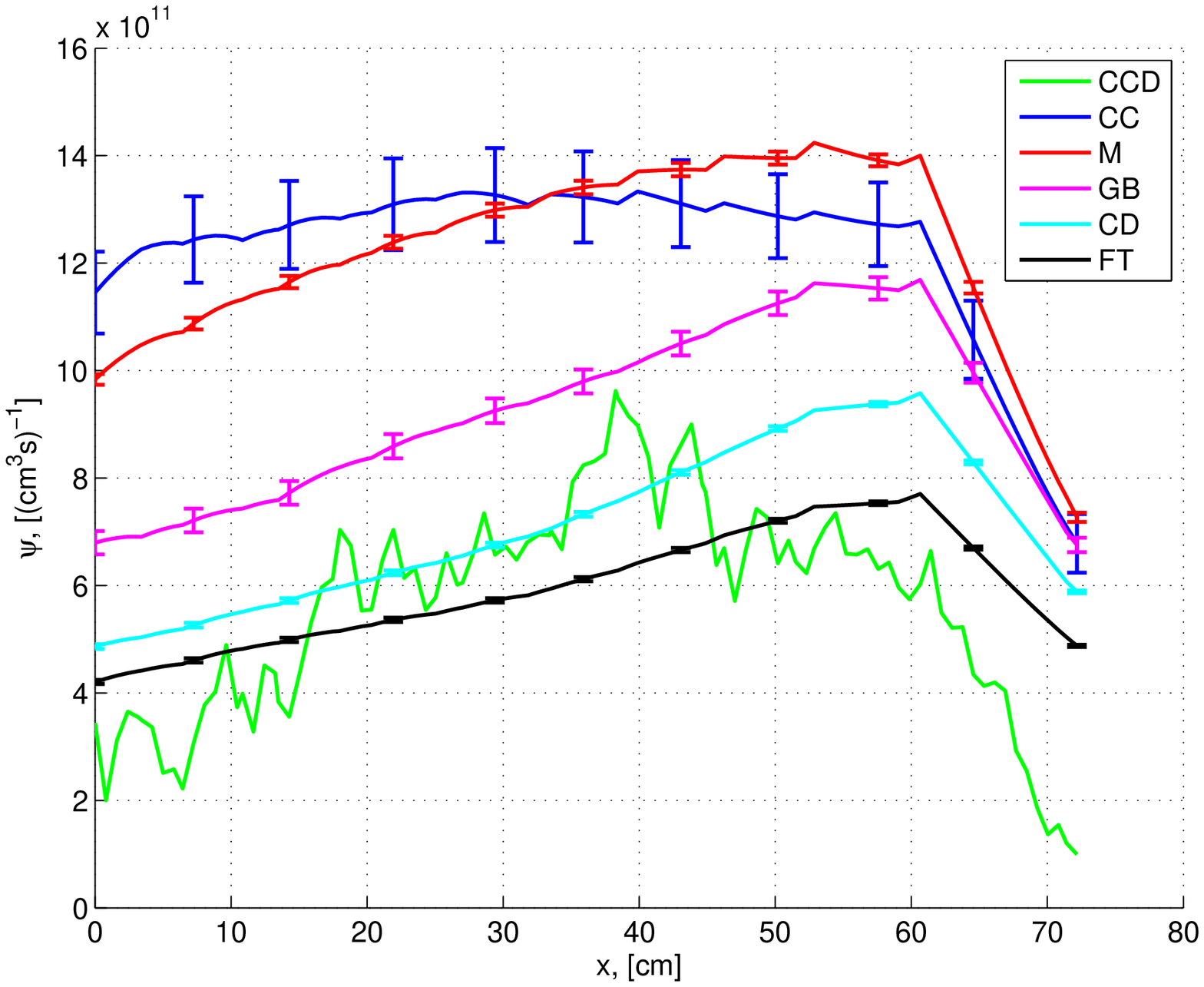}\\
\includegraphics[width=5.3cm]{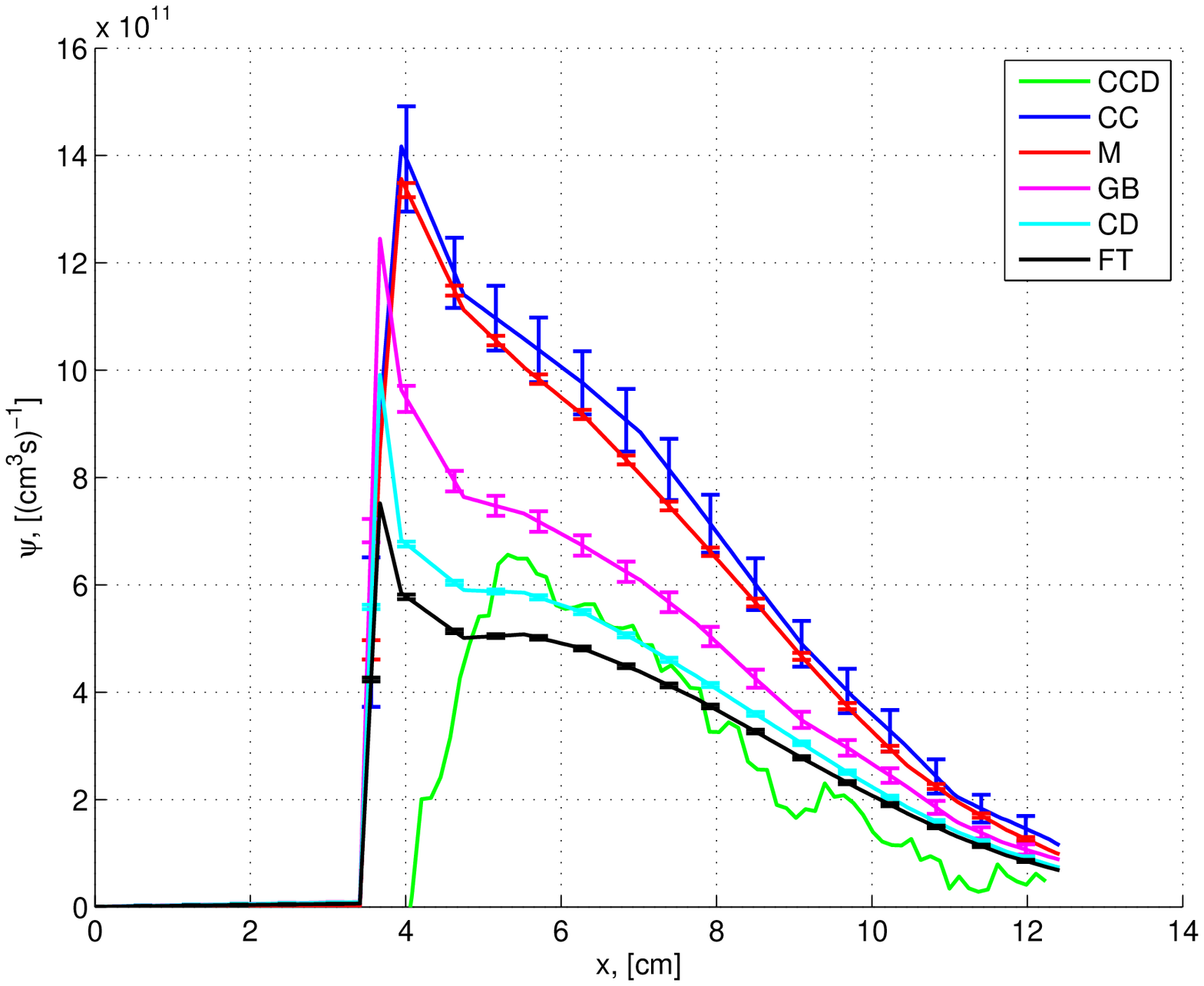}\\
\includegraphics[width=5.3cm]{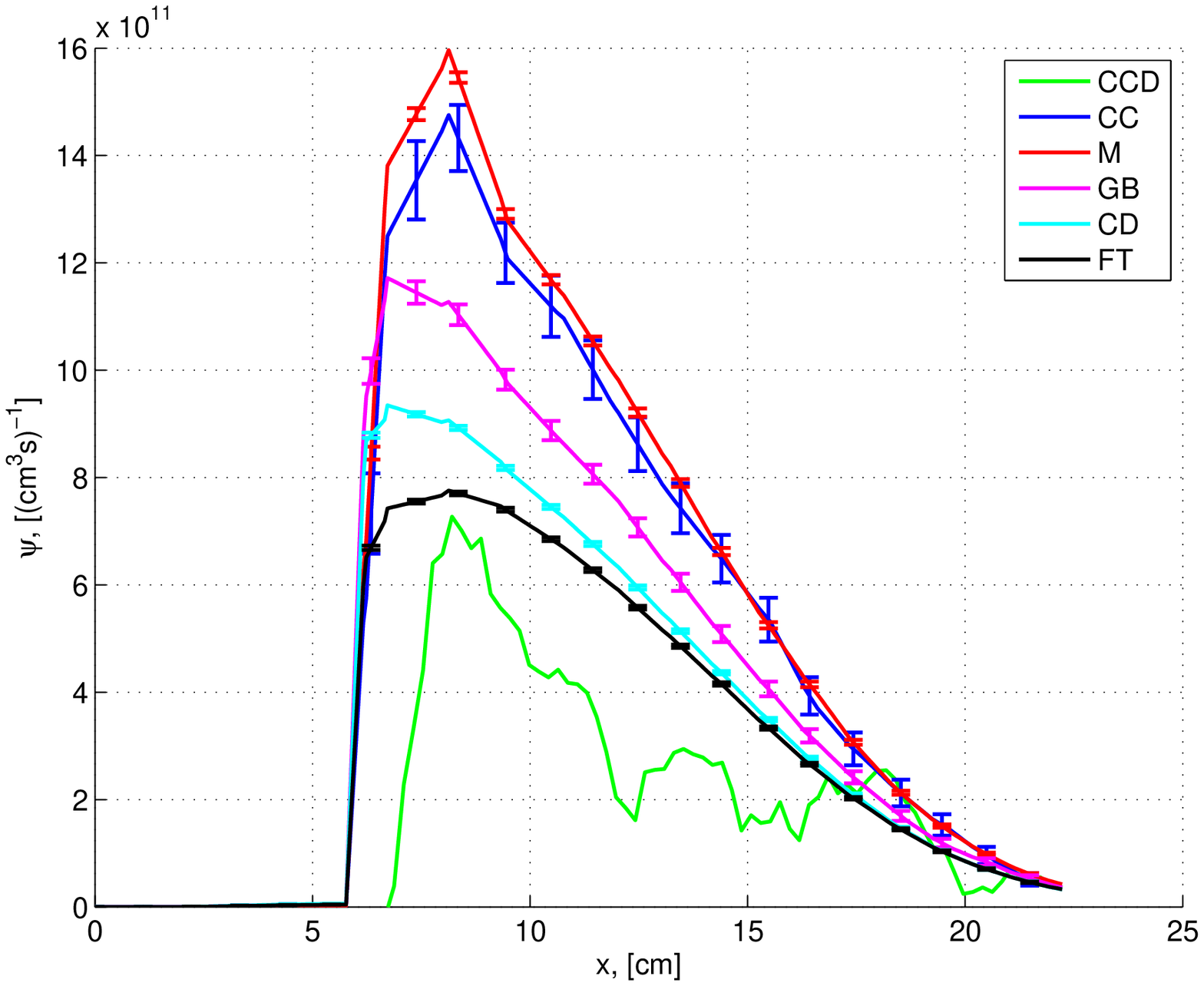}\\
\includegraphics[width=5.3cm]{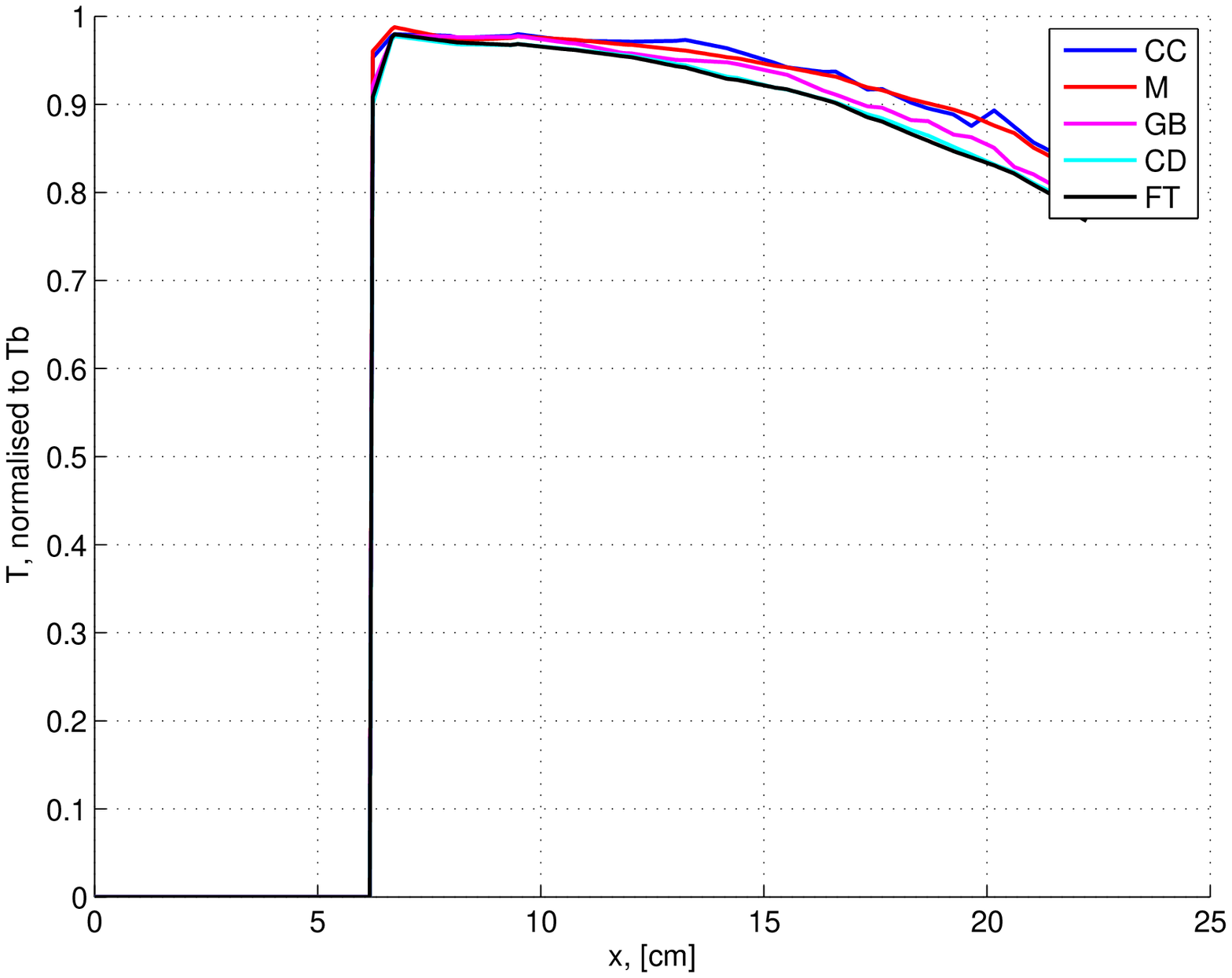}
\caption{CIII emission profiles from CCD camera (Mast lower divertor camera, by courtesy of S. Lisgo) vs. EIRENE simulation with \textit{trace ion module} including different effects; a.) cut 1 b.) cut 2 c.) cut 3 as indicated in figure 6b; d.) radial temperature profile, cut 3 in figure 6b}
\label{PIC_MAST_13949_CIII_Profiles_PSI}
\end{center}
\end{figure}
The extended EIRENE code supplemented by the \textit{trace ion module} has been tested for a realistic modeling case for the MAST tokamak \cite{MAST}. Due to the low aspect ratio of this tokamak $(A=R/a=1.3)$ neoclassical transport effects are enhanced, which makes MAST particularly well suited for testing the \textit{trace ion module}. The main purpose of the simulations was to see how different impurity transport processes affect impurity profiles, which has been supplemented by an at least qualitative comparison of the simulation results with CCD camera images. The present case is based on discharge \#$13949$, during which the CCD system recorded CII and CIII emission. The plasma background for this discharge has been provided by the OSM code \cite{LISG2}. The quality of the OSM background plasma calculation is important because the emission of CII and CIII is strongly dependent on the local plasma conditions. The $\nabla B$ drift is acting downward for positive ions. A toroidally symmetric methane gas puff has been simulated, thus a 2D simulation is sufficient. The puff location was lower inboard at $r=0.281\,m, y=-1.22\,m$. The fragmentation of the methane in the divertor plasma of MAST, where temperatures are about $\sim 1-20eV$, is characterized by a walk through of neutral as well as ionized molecules, which result into the final fragmentation products C and H (and their ions). The single code concept for both types of particles avoids rapid data transfer between codes, which would occur in the standard treatment where two kinetic codes are used. Moreover it is intrinsically ensured that atomic and molecular data is used consistently. In the present case the methane fragmentation has been described with the data obtained in \cite{LANG}. Ionization and recombination of carbon atoms/ions has been described with ADAS data \cite{Summ}, in particular \textit{adf11/scd94} and \textit{adf11/acd94}. The highest charge state of carbon included in the simulations was $C^{2+}$, because the recombination rate of $C^{3+}$ is too small for being a significant source of $C^{2+}$ under the plasma conditions at hand. Moreover the life time of $C^{3+}$ in the MAST divertor is already long enough for being completely thermalized and a fluid description is the better and more efficient choice. The grid which has been used is shown in figure \ref{PIC_MAST_13949_GRID}, the boundary of which has been specified as carbon. Physical and chemical sputtering has been turned off. 
\begin{figure*}[hbtp]
\begin{center}
\includegraphics[width=13cm]{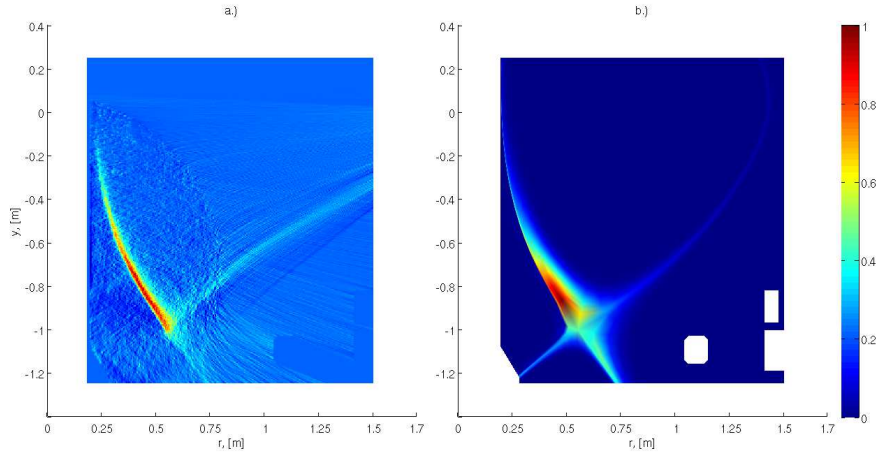}
\end{center}
\caption{a.) CII emission from CCD image, MAST discharge \#13949, $t=280ms$, (by courtesy of S. Lisgo) b.) CII emission from EIRENE simulation with \textit{trace ion module}}
\label{PIC_MAST_13949_CII}
\end{figure*}

Physics effects studied with the \textit{trace ion module} are friction force and thermalization, thermal force, mirror force, parallel electric field force and drift effects. Available drifts are $\nabla B$-drift, curvature drift including a quasi 3D correction term for 2D simulations and $E\times B$ drift, which requires an appropriate electric field model. If no effects for ions are activated particles simply follow magnetic field lines.

\begin{figure*}[hbtp]
\begin{center}
\includegraphics[width=13cm]{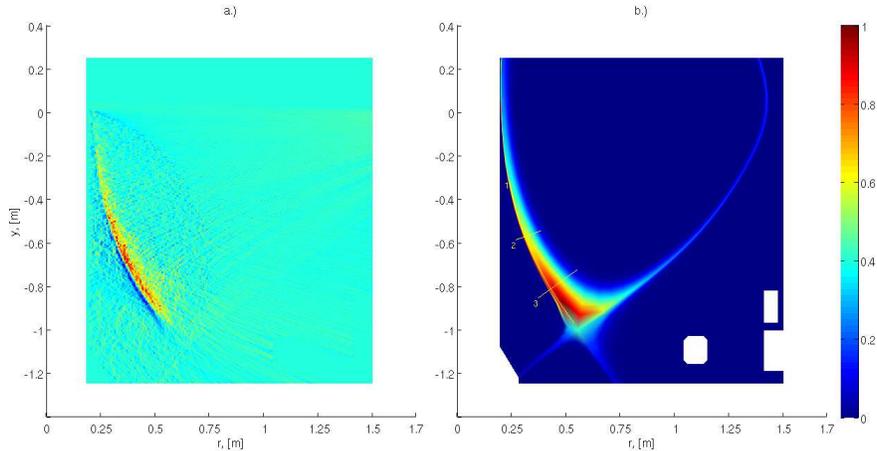}
\caption{a.) CIII emission from CCD image, MAST discharge \#13949, $t=280ms$, (by courtesy of S. Lisgo) b.) CIII emission from EIRENE simulation with \textit{trace ion module}}
\label{PIC_MAST_13949_CIII}
\end{center}
\end{figure*}

Radial and field aligned emission profiles with different effects active are shown in figure \ref{PIC_MAST_13949_CIII_Profiles_PSI} and have to be understood as indicated in the 2D emission pattern shown in figure \ref{PIC_MAST_13949_CIII} (two radial profiles 1,2 and one field aligned 3). Five different cases are shown: 1.) parallel motion along field lines and friction force $+$ thermalization (CC), 2.) mirror force added (M), 3.) $\nabla B$ drift added (GB), 4.) curvature drift added including quasi 3D correction (CD), 5.) $E\times B$ (FT) added. Profiles obtained from CCD camera data are denoted by CCD in the figures. The electric field has been obtained from the model described in Ref. \cite{Rei1}, the calculation of which only involves background plasma quantities. Because temperature gradients in the region of the radiated $CIII$ emission (inside the closed flux surface region) are small, the thermal force effect is negligible in this case and such simulations are not shown. Temperature profiles have been normalized to the background plasma temperature $T_b$, thus complete thermalization is reached at a value of $1$.

The general behavior of these emission profiles is a maximum near the separatrix and a decrease in opposite radial direction further in the plasma. Considering friction force and thermalization of $C^{2+}$ the profile is relatively flat along the field lines and falls off radially inward, following almost the inverse of the temperature profile (which mainly governs the lifetime and hence the density of the particles). The mirror force does not affect the radial profiles but particles are shifted further upstream to the inboard side along the field lines. The $\nabla B$ drift is acting in downward direction in this case and yields a decrease in density because particles are pushed to the bottom across the separatrix or in other words the probability for particles crossing the separatrix is lower. The major part of the curvature drift is due to toroidal curvature (for which a correction term has been introduced, see equ. \ref{CD_quasi3d_correction}) and is acting in direction of the $\nabla B$-drift. Thus it has a similar effect and further decreases the particle density. The last effect which has been added is an electric field model which slightly alters the density profiles to lower values. It has been observed that outside the separatrix the density profiles are practically zero, where it is noted that densities and temperatures are considered only in regions where standard deviation is below $10\%$. Whereas the emission profile is significantly influenced by the drifts the effect of the drifts on the temperature profile is small, see figure \ref{PIC_MAST_13949_CIII_Profiles_PSI}. In all cases $C^{2+}$ is nearly thermalized near the separatrix and the degree of thermalization decreases further into the plasma to $80\%$. Close to the separatrix the lifetime of $C^{2+}$ is about $200\,\mu s$, which is just long enough to fully thermalize. Further inside the plasma the lifetime is decreasing and full thermalization cannot be reached anymore. The observation from the simulation point of view is that the temperature behavior is exclusively governed by the Coulomb collision operator and the drift effects mainly act on the density profiles.

Finally the simulation results have been compared at least qualitatively with uncalibrated CCD camera images for carbon emission at two wave length ($465nm$, CIII and $514nm$, CII) from MAST discharge 13949, which have been kindly provided by S. Lisgo. It is noted that raw images of the CCD system (resolution $\sim5\,mm$) have been converted into 2D poloidal emissivity profiles via tomographic inversion and emission from intrinsic carbon due to sputtering has been subtracted. Simulated carbon density profiles have been converted into emission profiles using photon emissivity coefficients (PEC) according to 
\begin{equation}
    \psi = n_i\,n_e\,PEC(\lambda,n_e,T_e)
\end{equation}
where $n_i$ is the particle density of the particular species, $n_e$ and $T_e$ are the electron density and temperature and $\lambda$ is the specific wave length. The particular PEC's have been taken from the ADAS database ($pec96\#c\_vsu\#c1.dat$ and $pec96\#c\_vsu\#c2.dat$, metastable states are not resolved). Results are shown for CII emission in figure \ref{PIC_MAST_13949_CII} and for CIII in \ref{PIC_MAST_13949_CIII}. 
The general behavior of the simulated CII profile corresponds to the experimental one, but especially the density extends much further into the plasma as experimental seen. The CII emission has been calculated from the simulated $C^+$ density profile, which is governed mainly by the source of $C^+$. In this case the source of $C^+$ is determined by the break up process of methane, where finally the dissociation of $CH$ is the only channel to produce $C$ and subsequently $C^+$. In the present simulation the Langer database has been used to describe the break up of the methane, which might allow to penetrate $CH$ or $C$ too far into the plasma before being ionized to $C^+$. The influence of the present break up model has not been fully understood yet and will be further investigated. 

In contrast the lifetime of $C^{2+}$ is much longer than for $C^+$ and the density profile is more affected by the transport rather than the source profile. The radiation close to the X point is overestimated by the simulation, but the radiation pattern itself is close to the CCD image. This is explained by the low aspect ratio of MAST where the strong magnetic field curvature is significantly enhancing the drift effects, which govern the deposition of the particles. 

The comparison of the CCD images and the trace ion transport simulations with the extended EIRENE code show at least that the overall model is consistent with the experimental observations.

Finally a comment on the computational demands of the performed simulations is given. It has been observed that approximately $10^5$ particle histories are required for the present MAST case to obtain a reasonably small standard deviation in regions of interest ($<10\%$). Might be different for other cases. Tracing $10^5$ particles on the grid shown in figure \ref{PIC_MAST_13949_GRID}, which consists of $13536$ triangles, took approximately $10$ minutes on an Intel Core2 Quad at 2,83 Ghz. The maximum allowed time step for the tracing ions was $10^{-7}s$ and the maximum life time of a particle has been limited to $0.1s$. On the Opteron cluster LEO I of Innsbruck University an EIRENE run where exactly $10^6$ particle histories have been calculated on 96 cpu's took less than one minute. Leo I is a mixture of Opteron Barcelona at 2,3 Ghz and Opteron Shanghai at 2,5 Ghz. 
\section{Conclusions} \label{section_conclusions}
A linear kinetic Monte Carlo particle transport algorithm for
neutral (force-free) particles and linear Boltzmann collision
integrals has been supplemented by numerical orbit integration
modules as well as diffusive (Fokker-Planck) velocity space
collision integrals. The main motivation for this development was to
combine neutral atom transport with drift (guiding center) kinetic
impurity ion transport in magnetized fusion edge plasmas into a
single algorithm.

The unified procedure allows to simulate both neutral and charged
particles within a single source code and hence to avoid any
numerical dissipation which would otherwise be caused by frequent
transfer of information between neutral particle to charged particle
modules. In particular the extended code seems to be well suited to
deal with the rich chemistry of e.g. hydrocarbons in fusion edge
plasmas, when a fragmentation of a single molecule proceeds via a
sequence of neutral and charged states with quite distinct transport
characteristics. Also in other circumstances, often in those
relevant for cool dense fusion divertor plasmas, such as transport
of weakly ionized trace impurities near target surfaces, a strictly
consistent transport description of neutral and charged states of an
atom can be important.

Clearly the optimal coordinate systems for solving transport
problems for magnetized ions and for neutrals are at odds with
each other, with the former necessarily being
related to (inhomogeneous) magnetic field properties, and the latter
being simply Cartesian coordinates (in which also solid surfaces and
hence boundary conditions for the ion transport are naturally
formulated).

We have shown that a computationally affordable orbit integration scheme,
which can be readily implemented into Cartesian grid based Monte
Carlo solvers, can be constructed and is sufficiently accurate for a 
number of relevant applications including those mentioned above. 
The boundaries of the validity range, for acceptable time steps 
(and hence acceptable CPU costs of the entire package) 
are specified and found to accommodate in particular typical 
fusion edge plasma and divertor plasma conditions.

The linear guiding center Fokker-Planck collision integral is
integrated into the Monte Carlo procedure by utilizing Rosenbluth
potentials, evaluated ``on the fly'' on non-Maxwellian background
velocity distributions to accommodate kinetic thermal force effects.
We have shown that this is economically possible and that the
unavoidable singularity in these potentials at the zero of
perpendicular velocity can be properly dealt with by an implicit
procedure of Monte Carlo sampling of the post-collision ion
velocities.

A series of test cases based on semi-analytic solutions to strongly
simplified problems has been developed to verify both the
implementation of the new collision kernel and the explicit orbit
integrator.

As an illustrative sample of the essential capabilities we apply the
extended code to physical and configurational model parameters
chosen to simulate a methane gas puff experiment carried out in the
compact spherical tokamak MAST (UK). In this magnetic configuration
and plasma density range  kinetic ion drift orbit effects can be
expected to be particularly pronounced.

Experimental results on weakly ionized carbon ion emission patterns
are discussed with respect to relative importance of some of the new
features in the extended code.

\section{Acknowledgments}
This work was partially funded by FWF (Austrian Science Fund) under project P21061. 

The author thanks Prof. D. Reiter, Forschungszentrum J\"ulich (Germany), for his support and for numerous of fruitful discussions on kinetic Monte Carlo methods. Moreover S. Lisgo, formerly at MAST (Culham science center UK) is acknowledged for supplying OSM plasma backgrounds and CCD camera data for MAST. Felix Rei- mold, Forschungszentrum Garching (Germany), is acknowledged for extensive testing of the trace ion module for EIRENE. Finally the author thanks Assoc.-Prof. Siegbert Kuhn for his kind support during this work.

This work was supported by the Austrian Ministry of Science BMWF as part of the Uni Infrastrukturprogramm of the Forschungsplattform Scientific Computing at LFU Innsbruck. 

This work, supported by the European Communities under the Contract of Association between EURATOM and the Austrian Academy of Sciences, was carried out within the framework of the European Fusion Development Agreement. The views and opinions expressed herein do not necessarily reflect those of the European Commission.
\appendix
\section{Backward Orbit Integrator} \label{A_ADAMS_BASHFORTH}
The algorithm for advancing particle positions in the EIRENE code is given by
\begin{eqnarray}
 \textbf{y}_{t+\Delta t} = \textbf{y}_{t} + \dot{\textbf{y}}\cdot\Delta t,  \nonumber
\end{eqnarray}
where $\textbf{y}$ is the guiding center position in case of ionized particles and $\dot{\textbf{y}}$ is an effective guiding center velocity. Computationally affordable methods for evaluating $\dot{\textbf{y}}$, which fit to the Monte Carlo framework of EIRENE, are Adams Bashforth backward methods. The backward formula for an effective guiding center velocity for different numbers of backward steps $j$ is given by  
\begin{eqnarray}
 \dot{\textbf{y}}_{j} = \sum_{i=0}^{4}a_{ij}\cdot\dot{\textbf{y}}_{t-i\Delta t}, \nonumber
\end{eqnarray} 
where $a_{ij}$ contains coefficients corresponding to the number of backward steps required. If $j=1$ the first order Euler method is restored, where the effective guiding center velocity is calculated solely from the current plasma and magnetic field parameters at the current location as is described in detail in chapter \ref{subsection_langevin}. It is noted that $\dot{v}_{\parallel,\perp}$ have to be integrated in the same manner as $\dot{\textbf{y}}$. For a 4 step backward method the following coefficients have been calculated:
\begin{eqnarray}
a_{ij}=\left[\begin{array}{ccccc}
	1 & 0 & 0 & 0 & 0 \vspace{1mm}\\
	\frac{3}{2} & -\frac{1}{2} & 0 & 0 & 0 \vspace{1mm}\\	
	\frac{23}{12} & -\frac{16}{12} & \frac{5}{12} & 0 & 0 \vspace{1mm}\\	
	\frac{55}{24} & -\frac{59}{24} & \frac{37}{24} & -\frac{9}{24} & 0 \vspace{1mm}\\	
	\frac{1901}{720} & -\frac{1387}{360} & \frac{218}{60} & -\frac{637}{360} & \frac{251}{720}.
\end{array}\right]\nonumber 
\end{eqnarray}
Guiding center orbits are started with the Euler method and take each following time step on more precedent guiding center velocity, until four values are available. These backward methods require no additional computational effort, just one more array has to defined which stores the values of $\dot{\textbf{y}}$ and $\dot{v}_{\parallel,\perp}$ for $j$ successive steps. 
\section{Coulomb Collision Operator for Non Max- wellian Plasmas} \label{A_13MINT}
In the test particle approach the treatment of collisions with the background plasma (denoted by index $_b$), given by fluid quantities, requires the construction of a kinetic distribution function $f_b$. One approach, taking into account both the friction force as well as the thermo effect, is to describe the local plasma equilibrium by a Maxwellian like reference state, perturbed by a distortion function accounting for inhomogeneities. The distortion function can be a series of Laguerre Polynomials, e.g. \cite{SOLDOR}, or Hermite Polynomials, e.g. \cite{Rei1,Rei2}. In the present work the 13 moment approximation is considered, where the deviation from equilibrium $\eta(\textbf{v}_b,t)$ is given by a truncated series of Hermitian polynomials. This approach has been developed in \cite{Bal1} and applied to plasma transport codes in \cite{Rei1,Rei2}. The calculation of a kinetic distribution function with this method requires in addition to the five fluid moments number density, $n_b(\textbf{x})$, temperature, $T_b(\textbf{x})$ and flow velocity, $\textbf{u}_b(\textbf{x})$, three components of the heat flux $\textbf{q}(\textbf{x})$ and five independent components of the pressure tensor $\boldsymbol\pi$, hence 13 moments are used. The space coordinate is denoted by $\textbf{x}$. For deriving appropriate drift and diffusion coefficients in the standard Trubnikov/Rosenbluth formalism it is convent to use the inverse thermal velocity, $v_{th}=\sqrt{m_b/2T_b}$, as a normalization parameter, 
\begin{equation}
   \alpha=\frac{1}{v_{th}}=\sqrt{\frac{m_b}{2T_b}}, \label{THERMAL_VEL}
\end{equation}
and furthermore to introduce the following normalized velocities
\begin{equation}
	\textbf{c}=\alpha(\textbf{v}_b-\textbf{u}_b(\textbf{x}) ), \hspace{0.2cm} \bm{\chi} =\alpha (\textbf{v}-\textbf{u}_b(\textbf{x}) ),  \hspace{0.2cm} \textbf{y}=\textbf{c}-\bm{\chi}, \label{NORMALIZATIONS}
\end{equation}
where $v_b$ is the single particle velocity of background particles and $v$ denotes the test particle velocity. Then the perturbed Maxwellian distribution function is given by
\begin{equation}
    f_{b}(\textbf{x},\textbf{c}) = n_b(\textbf{x})\frac{\alpha^3}{\pi^{3/2}}\exp\left(-\textbf{c}^2\right)(1+\eta(\textbf{x},\textbf{c})).
\end{equation}
The distortion function in the 13 moment approximation reads (summation over i and j)
\begin{equation}
    \eta(\textbf{x},\textbf{c})=h_{i}^{(3)}(\textbf{x})H_i^{(3)}(\textbf{c})+h_{ij}^{(2)}(\textbf{x})H_{ij}^{(2)}(\textbf{c}),
\end{equation}
with the definition of the Hermitian polynomials in terms of normalized velocities,
\begin{eqnarray}
    H_i^{(3)}(\textbf{c})&=&\frac{1}{\sqrt{5}}c_i\left(2\textbf{c}^2-5\right) \nonumber\\
    H_{ij}^{(2)}(\textbf{c})&=&\sqrt{2}\left(c_i c_j-\frac{1}{3}\textbf{c}^2\delta_{ij}\right). \label{hermit_poly}
\end{eqnarray}
The corresponding coefficients are related to the heat flux $\textbf{q}(\textbf{x})$ and the pressure tensor $\boldsymbol\pi$ by
\begin{equation}
    h_{i}^{(3)}=\sqrt{\frac{2\,m_b}{5\,T_b^3}}\frac{1}{n_b}q_i \hspace{1cm}     h_{ij}^{(2)}=\frac{1}{\sqrt{2}}\frac{1}{n_b\,T_b}\pi_{ij}. \label{coeff_hermite}
\end{equation}
By substitution of $\textbf{c}=\textbf{y}+\bm{\chi}$ the perturbed Maxwellian can be written as 
\begin{eqnarray}
    f_b(\textbf{y}+\boldsymbol\chi) &=& \frac{n_b(\textbf{x})\alpha^3}{\pi^{3/2}}e^{-(\textbf{y}+\boldsymbol\chi)^2}\nonumber\\&& \left(1+\sum_i h_{i}^{(3)}H_i^{(2)}(\textbf{y}+\boldsymbol\chi)\right.\nonumber\\&& \left.+\sum_{i,j}h_{ij}^{(2)}H_{ij}^{(3)}(\textbf{y}+\boldsymbol\chi)\right).
\end{eqnarray}
and the Trubnikov/Rosenbluth potentials can be written as convolution integrals of the form
\begin{eqnarray}
    \phi(\textbf{x},\boldsymbol\chi)&=& \int_{\mathbb{R}^3} \frac{d^3y}{\alpha^3} \frac{\alpha}{y}f_b(\textbf{y}+\boldsymbol\chi) \nonumber\\
    \psi(\textbf{x},\boldsymbol\chi)&=& \int_{\mathbb{R}^3} \frac{d^3y}{\alpha^3} \frac{y}{\alpha}f_b(\textbf{y}+\boldsymbol\chi).
\end{eqnarray}
The potential functions can then be written as a sum of integrals depending on the normalized test particle velocity $\chi_i$,
\begin{eqnarray}
    \phi(\textbf{x},\boldsymbol\chi)&=&n_b(\textbf{x})\left(I^1(\boldsymbol\chi)+\right.\nonumber\\&&\left.\sum_ih_i^{(3)} I_i^3(\boldsymbol\chi)+\sum_{i,j}h_{ij}^{(2)}I^5_{ij}(\boldsymbol\chi)\right) \nonumber \\
    \psi(\textbf{x},\boldsymbol\chi)&=&n_b(\textbf{x})\left(I^2(\boldsymbol\chi)+\right.\nonumber\\&&\left.\sum_i h_i^{(3)} I_i^4(\boldsymbol\chi)+\sum_{i,j} h_{ij}^{(2)} I^6_{ij}(\boldsymbol\chi)\right). \label{POT_13M_6INT}
\end{eqnarray}
The integrals $I^{1-6}$ and the details of the evaluation as well as the corresponding results are given in \ref{potential_functions_integrals}. By introducing a generalized heat flux $\textbf{Q}$ and a generalized pressure $\boldsymbol\Pi$ tensor with
\begin{equation}
      Q_i=-\frac{1}{\sqrt{5}}h_{i}^{(3)} \hspace{1cm} \Pi_{ij}=\frac{1}{2\sqrt{2}}h_{ij}^{(2)},
\end{equation}
and skipping $n(\textbf{x})$ the Trubnikov/Rosenbluth potentials can be written as
\begin{eqnarray}
	\phi(\boldsymbol\chi)&=&\phi_0+\sum_i Q_i\frac{\chi_i}{\chi}\phi_1\nonumber\\&&+\sum_{ij}\Pi_{ij}\left(\delta_{ij}-3\frac{\chi_i\chi_j}{\chi^2}\right)\phi_2 \nonumber \\
        \psi(\boldsymbol\chi)&=&\psi_0+\sum_i Q_i\frac{\chi_i}{\chi}\psi_1\nonumber\\&&+\sum_{ij}\Pi_{ij}\left(\delta_{ij}-3\frac{\chi_i\chi_j}{\chi^2}\right)\psi_2 \label{POT_13M_FULL}
\end{eqnarray}
where it is noted that the potential functions $\phi_{0,1,2}$ and $\psi_{0,1,2}$, which are defined in \ref{POTENTIALS}, depend only on the absolute value of  $\chi=\parallel\boldsymbol\chi\parallel$ and are ab initio independent of the gyro angle.
\section{Potential Functions of the Coulomb Collision Operator} \label{potential_functions_integrals}
For obtaining the Trubnikov/Rosenbluth potentials for a perturbed Maxwellian velocity distribution function the following integrals have to be calculated
\begin{align}
    &I^1(\boldsymbol\chi)= \frac{1}{\pi^{3/2}}\int_{\mathbb{R}^3}{d^3y\,\frac{\alpha}{y}\, e^{-(\textbf{y}+\boldsymbol\chi)^2}}\nonumber \\
    &I^2(\boldsymbol\chi)= \frac{1}{\pi^{3/2}}\int_{\mathbb{R}^3}{d^3y\,\frac{y}{\alpha}\, e^{-(\textbf{y}+\boldsymbol\chi)^2}}\nonumber \\
    &I^3_i(\boldsymbol\chi)= \frac{1}{\pi^{3/2}}\int_{\mathbb{R}^3}{d^3y\,\frac{\alpha}{y}\, e^{-(\textbf{y}+\boldsymbol\chi)^2}}H_i^{(3)}(\textbf{y}+\boldsymbol\chi)\nonumber \\
    &I^4_i(\boldsymbol\chi)= \frac{1}{\pi^{3/2}}\int_{\mathbb{R}^3}{d^3y\,\frac{y}{\alpha}\, e^{-(\textbf{y}+\boldsymbol\chi)^2}}H_i^{(3)}(\textbf{y}+\boldsymbol\chi)\nonumber \\
    &I^5_{ij}(\boldsymbol\chi)= \frac{1}{\pi^{3/2}}\int_{\mathbb{R}^3}{d^3y\,\frac{\alpha}{y}\, e^{-(\textbf{y}+\boldsymbol\chi)^2}}H_{ij}^{(2)}(\textbf{y}+\boldsymbol\chi)\nonumber \\
    &I^6_{ij}(\boldsymbol\chi)= \frac{1}{\pi^{3/2}}\int_{\mathbb{R}^3}{d^3y\,\frac{y}{\alpha}\, e^{-(\textbf{y}+\boldsymbol\chi)^2}}H_{ij}^{(2)}(\textbf{y}+\boldsymbol\chi)\label{INT_13M_A}
\end{align}
with the Hermitian polynomials defined as in \ref{hermit_poly}. The integration can be performed by switching to a spherical coordinate system $(\textbf{e}_r,\textbf{e}_{\theta},\textbf{e}_{\varphi})$ with $\textbf{e}_r$  parallel to the vector $\boldsymbol{\chi}$, then $\boldsymbol{\chi}$ and $\textbf{y}$ have the form
\begin{equation}
    \boldsymbol\chi=\left(\begin{array}{c} 0 \\ 0 \\ \chi \end{array}\right)\hspace{1cm} \textbf{y}=\left(\begin{array}{l} y\,\sin\theta\,\cos\varphi \\ y\,\sin\theta\,\sin\varphi \\ y\,\cos\theta \end{array}\right)
\end{equation}
and the squared absolute value of the sum of these two vectors reads
\begin{equation}
    (\textbf{y}+\boldsymbol\chi)^2=\chi^2+y^2+2\chi y\,\cos\theta.
\end{equation}
In such a coordinate system integrals of the following form arise
\begin{equation}
    I(\boldsymbol\chi)= \frac{1}{\pi^{3/2}}\int_{0}^{\infty}\int_{0}^{\pi}\int_{0}^{2\pi}y^2\sin\theta dy\,d\theta d\,\varphi\,...
\end{equation}
the integration of which is straight forward. Since the solutions of the integrals consist of combination of the error function and it's derivative we introduce
\begin{equation}
    \Phi = erf(\chi) \hspace{1cm} \Phi' = \frac{2}{\sqrt{\pi}}e^{-\chi^2},
\end{equation}
and define the following potential functions in the same way as in \cite{Rei1}
\begin{eqnarray}
    \phi_0 &=& \frac{\alpha}{\chi}\Phi\nonumber \\
    \phi_1 &=& \alpha\chi\Phi'\nonumber \\
    \phi_2 &=& \alpha\left(\frac{2}{3}+\frac{1}{\chi^2}\right)\Phi'-\frac{\alpha}{\chi^3}\Phi\nonumber \\
    \psi_0 &=& \frac{1}{2\alpha}\Phi'+\frac{1}{\alpha}\left(\frac{1}{2\chi}+\chi\right)\Phi\nonumber \\
    \psi_1 &=& \frac{1}{2\alpha\chi}\Phi'-\frac{1}{2\alpha\chi^2}\Phi\nonumber \\
    \psi_2 &=& \frac{1}{2\alpha\chi^2}\Phi'-\frac{1}{\alpha}\left(\frac{1}{3\chi}-\frac{1}{2\chi^3}\right)\Phi. \label{POTENTIALS}
\end{eqnarray}
Using these definitions one can write the solutions of the integrals in a convenient way
\begin{equation}
   I^1(\boldsymbol\chi)=\phi_0 \hspace{1cm} I^2(\boldsymbol\chi)=\psi_0 \nonumber 
\end{equation}
\begin{equation}    
    I^3_i(\boldsymbol\chi)=-\frac{1}{\sqrt{5}}\phi_1\textbf{e}_r \hspace{1cm}    I^4_i(\boldsymbol\chi)=-\frac{1}{\sqrt{5}}\psi_1\textbf{e}_r \nonumber
\end{equation}
\begin{eqnarray} 
    I^5_{ik}(\boldsymbol\chi)=\left(\begin{array}{ccc}
    ×\frac{1}{2\sqrt{2}}\phi_2 & 0 & 0 \\
       0 & \frac{1}{2\sqrt{2}}\phi_2 & 0 \\
       0 & 0 & -\frac{1}{\sqrt{2}}\phi_2   
    \end{array}\right)\nonumber\\
    I^6_{ik}(\boldsymbol\chi)=\left(\begin{array}{ccc}
    ×\frac{1}{2\sqrt{2}}\psi_2 & 0 & 0 \\
       0 & \frac{1}{2\sqrt{2}}\psi_2 & 0 \\
       0 & 0 & -\frac{1}{\sqrt{2}}\psi_2   
    \end{array}\right)
\end{eqnarray}
These results have been calculated for a coordinate system with $\textbf{e}_r$ parallel to $\boldsymbol\chi$ one has to transform it back into the coordinate system of the laboratory, which can be done by a rotation defined by two matrices
\begin{eqnarray}
    \mathbb{R}_z(\varphi)&=&\left(\begin{array}{ccc}
    \cos\varphi & -\sin\varphi & 0 \\
       \sin\varphi & \cos\varphi & 0 \\
       0 & 0 &  0  
    \end{array}\right)
 \nonumber\\
    \mathbb{R}_y(\theta)&=&\left(\begin{array}{ccc}
    × \cos\theta & 0 & \sin\theta \\
       0 &  & 0 \\
       -\sin\theta & 0 & \cos\theta 
    \end{array}\right)
\end{eqnarray}
A vector in the coordinate system $(\textbf{e}_r,\textbf{e}_{\theta},\textbf{e}_{\varphi})$ can be transformed to the laboratory system $(\textbf{e}_x,\textbf{e}_y,\textbf{e}_z)$ denoted by $L$ via
\begin{equation}
   \textbf{I}^{3,4}_L=(\textbf{e}_x,\textbf{e}_y,\textbf{e}_z)\mathbb{R}_z(\varphi)\mathbb{R}_y(\theta)\left(\begin{array}{c} 0 \\ 0 \\ I_r^{3,4} \end{array}\right)
\end{equation}
and a matrix 
\begin{equation}
   I^{5,6}_L=MI^{5,6}M^{T} \hspace{0.5cm}  M=\mathbb{R}_z(\varphi)\mathbb{R}_y(\theta)
\end{equation}
Using the following relations of the angles and the coordinates in the laboratory system
\begin{equation}
    \frac{\chi_x}{\chi}=\sin\theta\cos\varphi \hspace{0.3cm} \frac{\chi_y}{\chi}=\sin\theta\sin\varphi \hspace{0.3cm}
    \frac{\chi_z}{\chi}=\cos\theta
\end{equation}
the final results of the integrals can be written as
\begin{eqnarray}
    I^1(\boldsymbol\chi)&=&\phi_0\nonumber \\
    I^2(\boldsymbol\chi)&=&\psi_0\nonumber \\
    I^3_i(\boldsymbol\chi)&=&-\frac{1}{\sqrt{5}}\frac{\chi_i}{\chi}\phi_1\nonumber \\
    I^4_i(\boldsymbol\chi)&=&-\frac{1}{\sqrt{5}}\frac{\chi_i}{\chi}\psi_1\nonumber \\
    I^5_{ij}(\boldsymbol\chi)&=&\frac{1}{2\sqrt{2}}\left(\delta_{ij}-3\frac{\chi_i\chi_j}{\chi^2}\right)\phi_2\nonumber \\
    I^6_{ij}(\boldsymbol\chi)&=&\frac{1}{2\sqrt{2}}\left(\delta_{ij}-3\frac{\chi_i\chi_j}{\chi^2}\right)\psi_2.\nonumber
\end{eqnarray}
These results coincide with the ones obtained by Reiser in \cite{Rei1} and \cite{Rei2}.
\section{Derivation of the Drift and Diffusion Coefficients}
The explicit form of the drift and diffusion coefficients \ref{diff_coeff_1} and the derivatives of the potential functions \ref{POT_13M_FULL} is given. In contrast to \cite{Rei1}, where derivatives have been calculated for the parallel direction only, also the perpendicular direction is considered in the present case, thus allowing to extend the thermo effect in perpendicular direction. This extension is planned in a future project. The following helpful quantities have been used,
\begin{align*}
	\frac{\partial \Phi_m}{\partial v_k}=\alpha\frac{\chi_k}{\chi}\Phi_m', \qquad \frac{\partial}{\partial \chi_i}\chi=\frac{\chi_i}{\chi},\\
        \frac{\partial}{\partial \chi_i}\frac{\chi_j}{\chi}=\frac{1}{\chi}\left(\delta_{ij}-\frac{\chi_i\chi_j}{\chi^2}\right)=N_{ij}.
\end{align*}
Then it follows for the first derivative with respect to $v_k$
\begin{align}
	&\frac{\partial \phi}{\partial v_k}=\,\alpha\frac{\chi_k}{\chi}\phi_0' \nonumber\\
        &\,+\,\alpha\sum_i Q_i\left[N_{ik}\phi_1+\frac{\chi_i\chi_k}{\chi^2}\phi_1'\right]\nonumber\\
        &\,+\,\alpha\sum_{ij}\Pi_{ij}\left[-3 M_{ikj}\phi_2+\left(\delta_{ij}-3\frac{\chi_i\chi_j}{\chi^2}\right)\frac{\chi_k}{\chi}\phi_2'\right] \label{first_deriv_pot} 
\end{align}
where 
\begin{equation}
M_{ikj}=\frac{\chi_i}{\chi}N_{kj}+N_{ik}\frac{\chi_j}{\chi}. \nonumber
\end{equation}
The second derivative reads
\begin{align}
	&\frac{\partial^2\phi}{\partial v_l\partial v_k}=\,\alpha^2\left[N_{lk}\phi_0'+\frac{\chi_k\chi_l}{\chi^2}\phi_0''\right]\nonumber\\
        &\,\,+\,\alpha^2\sum_i Q_i\bigg[\partial_l N_{ik}\phi_1+M_{ilk}\phi_1'\\
        &\hspace{1cm}+\frac{\chi_l}{\chi}\left(N_{ik}\phi_1'+\frac{\chi_i\chi_k}{\chi^2}\phi_1''\right)\bigg]\nonumber\\
        &\,\,+\,\alpha^2\sum_{ij}\Pi_{ij}\bigg[-3\partial_l M_{ikj}\phi_2\\
        &\hspace{1cm}-3\left(\frac{\chi_l}{\chi}M_{ikj}+M_{ilj}\frac{\chi_k}{\chi}\right)\phi_2'\nonumber\\
        &\hspace{1cm}+\left(\delta_{ij}-3\frac{\chi_i\chi_j}{\chi^2}\right)\left(N_{lk}\phi_2'+\frac{\chi_k\chi_l}{\chi^2}\phi_2''\right) \bigg] \label{second_deriv_pot}
\end{align}
where the following tensors have been introduced as
\begin{equation*}
	\partial_l N_{ik}=-\frac{\chi_l}{\chi^3}\left(\delta_{ik}-\frac{\chi_i\chi_k}{\chi^2}\right)-\frac{1}{\chi}M_{ilj}
\end{equation*}
\begin{equation*}
        \partial_l M_{ikj}=N_{li}N_{kj}+\frac{\chi_i}{\chi}\partial_l N_{kj}+\frac{\chi_j}{\chi}\partial_l N_{ik}+N_{ik} N_{lj}.
\end{equation*}
In addition to the derivatives of the potential functions \ref{POT_13M_FULL} the scalar product
\begin{equation}
     C1=\textbf{Q}\cdot\boldsymbol\chi \label{C1_dotprod}
\end{equation}
and the sum over all elements of the matrix
\begin{equation}
     C2=\sum_{ij}\Pi_{ij}\left(\delta_{ij}-3\frac{\chi_i\chi_j}{\chi^2}\right)
\end{equation}
have to be evaluated for obtaining the explicit form of the drift and diffusion coefficients. The resulting scalar quantities $C1$ and $C2$ are not independent of the gyro angle ab initio. Hence they do not fit into the drift kinetic model which is the aim of the present work. Neglecting perpendicular heat fluxes (same approach as in \cite{Rei1}) the scalar product \ref{C1_dotprod} is gyro angle free and can be evaluated by 
\begin{equation}
     C1=-\frac{1}{\sqrt{5}}\sqrt{\frac{2m_b}{5T_b^{3}}}\frac{1}{n_b}\chi_{\parallel}q_{\parallel}
\end{equation}
where the parallel heat flux is defined as
\begin{equation}
	q_{\parallel}=\frac{15}{8\sqrt{2\pi}}\frac{T_b^{5/2}}{m_b^{1/2}}\frac{16\pi^2\epsilon_0^2}{Z^4e^4n_b\lambda}\tilde{\kappa}_{\parallel}.
\end{equation} 
The value of the transport coefficient $\tilde{\kappa}_{\parallel}$ is $1/0.5657$ according to \cite{Bal1}. With this approach at least the thermo effect in parallel direction can be studied. According to \cite{Bal1} this viscosity tensor is invariant under rotations around $\textbf{b}$ and it is assumed that the scalar quantity $C2$ is ab initio independent of the gyro angle for magnetized plasmas. The reduced form of the viscosity tensor is given by
\begin{equation}
    \pi_{ik}=-\frac{12}{\sqrt{2\pi}}\frac{\pi^2\epsilon_0^2}{Z_b^4e^4\lambda}m_b^{1/2}T_b^{5/2}\eta_{\parallel}\nu_{ik}^{\parallel} 
\end{equation}
where
\begin{equation}
    \nu^{\parallel}=\frac{1}{2}\left( \begin{array}{ccc} 2\nu_{xx} & 0 & 0 \\
                                                             0 & \nu_{yy}+\nu_{zz} & 0\\
                                                           0 & 0 & \nu_{yy}+\nu_{zz} \end{array}
                                                           \right) \nonumber
\end{equation}
Here the x direction coincides with the direction of the magnetic
field vector. The explicit form of the components $\nu_{ii}$ is
\begin{equation}
    \nu_{ii}=2\frac{\partial v_{bi}}{\partial
    x_i}-\frac{1}{3}\nabla\textbf{v}_b\nonumber
\end{equation}
with the factor $\eta_{\parallel}=1.0/(1.2+0.8485Z_b^{-1})$.
\section{Diffusion Coefficients for Langevin Equations} \label{A_DRIFTDIFFCOEFF}
The Langevin approach for solving Fokker-Planck equations requires a decomposition of the diffusion tensor according to $D=BB^T$. The choice of $B$ is not unique which reflects somehow the fact that there is more information in a random walk trajectory than needed to solve the Fokker-Planck equation \cite{Kloe}. In this chapter a reasonable method is given which treats the case at hand where the diffusion tensor is a positive definite symmetric 2x2 matrix. It can be decomposed as
\begin{equation}
      D=T\Lambda T^T,
\end{equation}
where $T$ is an orthogonal matrix. Then the diagonal matrix $\Lambda$ contains the eigenvalues of $D$. Introducing another matrix $S$ with $\Lambda=SS^T$ it follows
\begin{equation}
      D=T\Lambda T^T=TSS^TT^T=TS(TS)^T=BB^T.
\end{equation}
Thus a choice for the matrix $B$ is given by $B=TS$, where $S$ contains the square roots of the eigenvalues of $D$, which explicitly calculated read
\begin{equation}
      \lambda_{\pm}=\frac{1}{2}(D_{11}+D_{22})\pm\frac{1}{2}\sqrt{(D_{11}-D_{22})^2+4D_{12}}.
\end{equation} 
The matrix $T$ can be calculated as
\begin{equation}
      T = \frac{1}{N} \left( \begin{array}{cc}
          -D_{12} & D_{22}-\lambda_- \\
          D_{11}-\lambda_+ & -D_{12}
      \end{array} \right)
\end{equation} 
where the norm of the eigenvectors is 
\begin{equation*}
N=\sqrt{D_{12}+(D_{11}-\lambda_+)^2}=\sqrt{D_{12}+(D_{22}-\lambda_-)^2}.
\end{equation*}
The explicit form of the matrix $B$ is then given by
\begin{equation}
      B = \frac{1}{N} \left( \begin{array}{cc}
          -D_{12}\sqrt{\lambda_+} & (D_{22}-\lambda_-)\sqrt{\lambda_-} \\
          (D_{11}-\lambda_+)\sqrt{\lambda_+} & -D_{12}\sqrt{\lambda_-}
      \end{array} \right). \label{BMATRIX}
\end{equation} 
This method has been implemented into the EIRENE code for determination of the elements of $B$ at each time step.
\section{Semi Implicit Method for Coulomb Collisions of Low Energetic Particles} \label{A_IMPL_METH}
Evaluating drift and diffusion coefficients \ref{diff_coeff_1} for purely Maxwellian distribution functions of the plasma background in the cylindrical coordinate system at hand yields a singularity in the perpendicular drift coefficient. Thus the present Monte Carlo procedure has to be supplemented by an implicit method for properly advancing discrete perpendicular velocity increments in time. The explicit form of the perpendicular drift coefficient written in terms of the potential function $\phi$ only reads
\begin{equation}
  A_{2}=\eta\left[\frac{\hat{\phi}^{'}}{\chi}\left(\mu\chi_{2}+\frac{1}{4}\frac{1}{\chi_{2}}\right)+\frac{1}{2}\frac{1}{\chi_{2}}\hat{\phi}\right], \label{KPER_OPTIMISED}
\end{equation} 
where $\hat{\phi}=\phi_0/\alpha$ and $\chi_1=0$ without loss of generality, thus $\chi=\sqrt{\chi_1^2+\chi_2^2}\equiv\chi_2$. Its singular behavior near $0$ is obvious from the Taylor series expansion,
\begin{equation}
	A_{2}(\chi\rightarrow 0)\sim\frac{2\alpha}{3\sqrt{\pi}}\frac{1}{\chi}+O(\chi)^2. 
\end{equation}
The normalized perpendicular velocity is formally advanced in one time according to
\begin{equation}
  \chi_{2}^{post-col}=\chi_{2}^{pre-col}+\int_t^{t+\Delta t}A_{2}(\chi_{2}(t'),\chi_{1})dt',
\end{equation}
which has been solved implicitly with the first order Euler method given by 
\begin{equation}
  \chi_{2}^{post-col}=\chi_{2}^{pre-col}+\Delta t\,A_{2}(\chi_{2}^{post-col}),\chi_{1}). \label{IMPL_EULER}
\end{equation}
For performing the iterations the derivative of $A_{2}$ with respect to $v_{2}$ is required, 
\begin{equation}
  \frac{\partial A_{2}}{\partial v_{2}} =\eta\alpha\left(a\,\frac{\hat{\phi}^{'}}{\chi}-b\,\hat{\phi}\right),
\end{equation}
where the coefficients $a$ and $b$ are defined as follows
\begin{align}
  &a=\mu\left(1-3\frac{\chi_{2}^2}{\chi^2}-2\chi_{2}^2\right)-\frac{1}{4}\left(\frac{1}{\chi_{2}^2}+\frac{3}{\chi^2}\right) \nonumber\\
  &b=2\mu\frac{\chi_{2}^2}{\chi^2}+\frac{1}{2}\left(\frac{1}{\chi_{2}^2}+\frac{1}{\chi^2}\right). 
\end{align}
The second order Newton Raphson method is sufficient to solve the actual system which is given by
\begin{align}
     &f(\chi_{2}^{post-col})=\chi_{2}^{post-col}-\chi_{2}^{pre-col}-\Delta t\,A_{2}(\chi_{2}^{post-col}) \nonumber\\
     &f'(\chi_{2}^{post-col})=1-\Delta t\,A_{2}'(\chi_{2}^{post-col}),
\end{align}
and the well known iteration formulae:
\begin{equation}
     x_{n+1}=x_{n}+\frac{f(x_n)}{f'(x_n)}
\end{equation}
The implicitly obtained result for $\chi_{2}^{post-col}$ is used to calculate an effective $A_{2}=(\chi_{2}^{post-col}-\chi_{2}^{pre-col})/\Delta t$ which then overwrites $A_{2}$ from the explicit scheme. 

In practice it is sufficient to activate the presented implicit method only in case of small values $\chi_2\ll 1$, because the extremely stiff behavior of $A_2$ near $0$ changes into a slightly increasing function for more positive values of $\chi_2$. A threshold for $\chi_2$ can be specified in the input of the code. This parameter has been set to $0.01$ (dimensionless) for performing trace carbon simulations in the MAST divertor region. Actually this number depends strongly on the mass difference of the collision partners. If the test particle is much heavier as the field particle this number should be increased and vice versa. 

Moreover $f(\chi_{\perp}^{post-col})$ is a monotonically increasing function without minima and maxima and less than $10$ iterations are required in almost any case to reach a sufficient accuracy. The iteration is currently stopped if the error of the normalized quantity $\chi_{2}$, which is of the order unity, drops below $10^{-6}$. The additional computational costs are raised by $10\%$ once the implicit method is activated, but only in this case Coulomb collisions are properly treated. 


\begin{thebibliography}{99}
%
\bibitem{Schn} R Schneider, Contrib. Plasma Phys., Vol. 46, 2006
%
\bibitem{EDGE2D} S. Wiesen, http://www.eirene.de/e2deir\_report\_30jun06.pdf, 2006
%
\bibitem{UEDGE} T Rognlien et al, Contrib. Plasma Phys., Vol. 34, 1994
%
\bibitem{DEGAS} D Stotler, C Karney, Contrib. Plasma Phys.,  Vol. 34, 1994
%
\bibitem{SOLDOR} K Shimizu et al, Nuclear Fusion, Vol. 49, 2009
%
\bibitem{Stan} P Stangeby, The Plasma Boundary of Magnetic Fusion Devices (Institute of Physics Publishing, 2000)
%
\bibitem{Sipi_Phd} S Sipila, Monte Carlo Simulation of charged particle orbits in the presence of radio frequency waves in tokamak plasmas  (PhD Thesis, Helsinki University of Technology, Finland, 1997)
%
\bibitem{Sipi} S Sipila et al, Contrib. Plasma Phys., Vol. 42, 2002
%
\bibitem{Rei1} D Reiser, Berichte des Forschungszentrums J\"{u}lich 3508, 1998
%
\bibitem{Rei2} D Reiser, Nuclear Fusion, Vol. 38, 1998
%
\bibitem{Reit} D Reiter, Fusion Science and Technology, Vol. 47, 2005
%
\bibitem{EIRE} EIRENE User Manual at www.eirene.de
%
\bibitem{Sven_Phd} S Wiesen, Nichtlineare Simulation von Photonentransport in Plasmen (PhD Thesis, Ruhr Universit\"at Bochum, 2005)
%
\bibitem{ECK2}  R Behrisch and W Eckstein, Sputtering by Particle Bombardment: Experiments and Computer Calculations from Threshold to MeV Energies (Springer 2007)
%
\bibitem{Summ} H Summers et al, Plasma Phys. Control. Fusion 44, 2002
%
\bibitem{Boern} P B\"orner et al, Light Sources 2004, IoP Conf. Series *182,* 407, 2004
%
%
%
\bibitem{CMPP} H Fehske et al, Computational Many-Particle Physics (Lecture Notes in Physics, Springer 2008)
%
\bibitem{Hela} P Helander and D J Sigmar, Collisional Transport in Magnetized Plasmas (Cambridge University Press 2002)
%
\bibitem{Kloe} P E Kloeden and E Platen, Numerical Solution of Stochastic Differential Equations (Springer 1992)
%
\bibitem{Bal1} R Balescu, Transport Processes in Plasmas (Elvesier Science Publishers 1988)
%
\bibitem{LANG} W Langer, Nuclear Fusion, Vol. 22, 1982
%
\bibitem{LISG2} S Lisgo et al, Journal of Nucl. Mat., Vol. 337, 2005
%
\bibitem{Meth} R K Janev, D Reiter, Phys. Plasmas 9, 4071, 2002
%
\bibitem{Etha} R K Janev, D Reiter, Phys. Plasmas 11, 780, 2004
%
\bibitem{Trub} B A Trubnikov, Reviews of plasma physics, Vol. 1, 1965
%
\bibitem{ZAIT} F S Zaitsev, Aktual'nye Voprosky Prikladnoi Metamatiki, pp. 80-87, 1989. Translated 1991 Plenum Publishing Coorporation
%
\bibitem{MAST} B Lloyd et al, Plasma Phys. Control. Fusion, Vol. 46, 2004 
%
\end{thebibliography}
\end{document}